\documentclass[11pt]{article}

\parskip \medskipamount
\usepackage[T1]{fontenc}
\usepackage{titling}
\usepackage{authblk}
\usepackage{amsmath,amssymb,amsthm,mathrsfs}
\usepackage{indentfirst}
\usepackage{subdepth}
\usepackage{graphicx}
\usepackage{pdfpages}
\usepackage{subfigure}
\usepackage{epstopdf}
\usepackage{url}
\usepackage{float}
\usepackage{appendix}

\usepackage{graphicx}
\usepackage{amsfonts}
\usepackage{amsmath}
\usepackage{amssymb}
\usepackage{url}
\usepackage{fancyhdr}
\usepackage{indentfirst}
\usepackage{enumerate}
\usepackage{amsthm}
\usepackage{color}
\definecolor{r}{rgb}{0,0,0}
\usepackage{dsfont}

\newcommand{\Var}{\mathrm{Var}}

\newcommand{\p}{\mathbb{P}}
\newcommand{\E}{\mathbb{E}}
\newcommand{\R}{\mathbb{R}}

\usepackage{etoolbox}
\usepackage{microtype}
\usepackage[plainpages=false,hyperfootnotes=false,colorlinks=true,citecolor=blue]{hyperref}
\usepackage[centerlast]{caption}

\usepackage{apacite}
\usepackage[round, sort]{natbib}

\title{\textbf{The optimal reinsurance strategy with
price-competition between two reinsurers}}
\date{\vspace{-6ex}}

\author{Liyuan Lin\thanks{Department of Statistics \& Actuarial Science, University of Waterloo, Canada, Email: l89lin@uwaterloo.ca},~~
Fangda Liu\thanks{Department of Statistics \& Actuarial Science, University of Waterloo, Canada, Email: fangda.liu@uwaterloo.ca},~~
Jingzhen Liu\thanks{Corresponding author, China Institute for Actuarial Science,  Central University of Finance and Economics, Email: janejz.liu@hotmail.com}, ~~and~
Luyang Yu \thanks{School of insurance, Central University of Finance and Economics, Email: yuluyang0413@hotmail.com\\
\rm\small{This work was supported by the National Natural Science Foundation of China (Grant No. 11771466, No. 11601540), and the Natural Sciences and Engineering Research Council of Canada (RGPIN-2020-04717, DGECR-2020-00340)}}}

\hypersetup{
  colorlinks=true,
  urlcolor=blue,
  linkcolor=red,
  citecolor= blue
}

\theoremstyle{definition}
\newtheorem{definition}{Definition}[section]

\theoremstyle{plain}
\newtheorem{proposition}{Proposition}[section]
\newtheorem{lemma}{Lemma}[section]
\newtheorem{theorem}{Theorem}[section]
\renewenvironment{proof}{\noindent {\bf Proof.}}{\hfill $\Box$}    \def\d{\,\mathrm{d}}
\def\sd{\mathrm{d}}

\begin{document}

\maketitle

\begin{center}
\textbf{\large Abstract}
\end{center}
We study optimal reinsurance in the framework of stochastic game theory, in which there is an insurer and two reinsurers. \textcolor{r}{A Stackelberg model is established to analyse the non-cooperative relationship between the insurer and reinsurers, where the insurer is considered as the follower and the reinsurers are considered as the leaders. The insurer is a price taker who determines reinsurance demand in the reinsurance market, while the reinsurers can price the reinsurance treaties. Our contribution is to use a Nash game to describe the price-competition between two reinsurers.} We assume that one of the reinsurers adopts the variance premium principle and the other adopts the expected value premium principle. The insurer and the reinsurers aim to maximize their respective mean-variance cost functions which lead to a time-inconsistency control problem. To overcome the time-inconsistency issue in the game, we formulate the optimization problem of each player as an embedded game and solve it via a corresponding extended Hamilton-Jacobi-Bellman equation. We find that the insurer will sign propositional and excess loss reinsurance strategies with reinsurer 1 and reinsurer 2, respectively.  When the claim size follows exponential distribution, there exists a unique equilibrium reinsurance premium strategy. Our numerical analysis verifies the impact of claim size, risk aversion and interest rates of the insurer and reinsurers on equilibrium reinsurance strategy and premium strategy, which can help to understand competition in the reinsurance market.\\

\noindent {\bf JEL:} C720, D210, G220.

\vspace{3mm}

\section{Introduction}

Reinsurance is an important tool for insurers to transfer risk and improve underwriting capabilities. Research focused on the issue of optimal insurance/reinsurance can be traced back to \cite{borch1960an}, in which the author showed that the stop-loss reinsurance is the optimal under the criterion of minimizing the variance of an insurer's retained loss and the reinsurance premium is calculated according to the expected value principle.
In \cite{Arrow1963Uncertainty}, the author obtained the same result by maximizing the expected utility of the final wealth of a risk-averse insurer. Thereafter, many scholars began to study this problem under different assumptions on  risk process,  reinsurance premium principle,  objective function, etc. One may refer to relevant works including \cite{Bai2008}, \cite{kaluszka2004},  and \cite{Cai2007Optimal}. It should be pointed out that these studies mainly focused on the a market with  only one insurer and one reinsurer. 
However, it is common in practice that an insurer can transfer risk to multiple reinsurance companies. Since reinsurers choose different premium principles and have different risk tolerances, they participate in the reinsurance treaty with different coverage levels. Thus, some researchers began to study the optimal reinsurance plan in the presence of multiple reinsurers. In a static one-period risk modelling setup, there are plenty of studies investigating optimal reinsurance problems between one insurer and two reinsurers by considering different reinsurance price principles and criterion of optimization (see \cite{Asimit2013Optimal}, \cite{Chi2014}, \cite{Cong2014}, \cite{boonen2016a} and \cite{BOONEN2020}). Stochastic control theory has also been widely used to solve such optimal reinsurance problems. For example, \cite{MENG2016182} studied an optimal reinsurance problem for an insurer, who aims to minimize the probability of ruin by partially transferring the insurable risk to two reinsurers.

It is worth noting that the aforementioned works adopt only the insurer's perspective. The reinsurance premium was treated as a fixed constant or a predetermined function. However, reinsurers can also adjust their reinsurance premium according to the insurer's reinsurance plan.
 The interaction between insurers and reinsurers is intensively discussed in the literature by using game theory, and it can be described as either a cooperative game or a non-cooperative game.
In the cooperative games category, \cite{Borch1960Reciprocal} studied optimal reinsurance contracts within the context of bargaining games and found the Nash bargaining solution. Thereafter, \cite{J1993Cooperative}, \cite{aase2009}, \cite{boonen2016}, and \cite{bailihua2017} followed this direction and studied various reinsurance problems. In the non-cooperative games category, the Stackelberg model is an important method to describe the relationship between insurers and reinsurers.
It can be dated back to \cite{stackelberg1934}, in which the authors formulated a strategic game with the leader firm moving first and the follower firm moving second. \cite{Gerber1984Chains} used the notion of Stackelberg equilibria to study chains of reinsurance in a static framework.
In the Stackelberg game established in \cite{Morozov1998A},  the insurer is the follower and chooses the loss-ratio limit, and the reinsurer plays the role of leader and chooses the cost of the reinsurance policy.
\cite{Lv2019Stochastic} studied the stochastic Stackelberg differential reinsurance game with the reinsurer as a leader and the insurer as a follower, and  obtained time-consistent equilibrium strategies under the time-inconsistent mean-variance framework.
In \cite{baiyanfei2020}, the authors incorporated  the unequal status between the insurer and reinsurer due to the issues of asymmetric information into a Stackelberg stochastic differential reinsurance-investment game problem.

To the best of our knowledge, the discussion on  multiple reinsurers and potential competition between each other, which are essential issues in practice, are missing in the literature.
To fill up this gap, in this paper, we establish a  Stackelberg model between one insurer and two reinsurers who compete on price, and study the optimal reinsurance policy in a market equilibrium.
We adopt the continuous-time mean-variance criterion, which incorporates the trade-off between risk and return for each company.
The mean-variance analysis developed in  \cite{Markowitz1952Portfolio} provides a fundamental basis for portfolio construction in a single period, and has long been recognized as the cornerstone of modern portfolio theory.
After Markowitz's pioneering work, the mean-variance model was soon extended to multiperiod portfolio selection; see, for example, \cite{Paul2011Lifetime}, \cite{Nils1971Multi}, and \cite{BERNARD1991An}. Since the mean-variance model exists as a non-linear function of the expected value of terminal wealth, the Bellman optimality principle is not applicable for dynamic mean-variance problems, and the time-inconsistency issue rises. Our paper faces the same  time-inconsistency issue. The mainstream method of dealing with time-inconsistent problems was proposed by \cite{RMyopia}. Thereafter, \cite{Basak2009Dynamic} solved the dynamic mean-variance portfolio problem and derived its time-consistent solution using dynamic programming. \cite{Bj2014A} and \cite{tomas2017} developed a theory for a class of time-inconsistent stochastic control problems in discrete-time and continuous-time models. They studied these problems by viewing them within the game-theoretic framework and seeking subgame perfect Nash equilibrium points.
The presenting paper follows this approach. Precisely, we consider three embedded games for the insurer and two reinsurers, and each of the three embedded games is played by the future incarnations of the players. Then, we obtain three systems of extended Hamilton-Jacobi-Bellman (HJB) equations and find the equilibrium strategies and equilibrium value function.

In this paper, the relationship between the insurer and reinsurers in the Stackelberg model is hierarchical. The reinsurers, as the leaders in this model, have the right to determine the reinsurance premium. \textcolor{r}{The representative insurer, as the follower, can only accept the reinsurance premium requested by reinsurers and determine the loss ceded to the reinsurers.} In addition, we assume that the two reinsurers apply different premium principles, which are the variance premium principle and expected value premium, to stand for different kinds of reinsurance companies and consider their competition on price, which has never been considered in the prior literature. Thus, we first find the equilibrium reinsurance strategies for the insurer under the premiums given by the two reinsurers. Then, we consider the price competition between two reinsurers when the insurer applies its equilibrium reinsurance strategy, which can be explained by the model proposed by \cite{Bertrand1883Th}. For each reinsurer, we obtain the equilibrium premium strategy when the other reinsurer's premium is given and then find the equilibrium strategy for price competition. Finally, by substituting the strategies of the two reinsurers into the insurer's reflection function, we obtain the equilibrium reinsurance strategies of the insurer. Our conclusion can lay a foundation for research about extension to multiple reinsurance companies.

The remainder of the paper is organized as follows. Section \ref{model} establishes the model dynamics. In section \ref{time-consitent}, we address the time-inconsistency issue and obtain the extended HJB equation. We solve the extended HJB equation of the insurer and obtain the equilibrium reinsurance strategy of the insurer. In section \ref{example}, we present an example and find the unique equilibrium reinsurance price strategy for two reinsurers under exponential claims. Section \ref{numerical} furthers our analysis numerically when given exponentially distributed claims. Section \ref{conclusion} concludes the paper. All technical proofs are relegated to Appendix \ref{appendix}.

\section{The Model}\label{model}

Consider an insurance market including one primary insurer and two reinsurers.
We use a complete filtered probability space $(\Omega,\mathcal{F},\{\mathcal{F}_t\}_{t\in[0,T]},\p)$ to model the insurance market, where $ T < \infty $ is the terminal time of decision making, $\p$ is a real-world probability measure, and $\{\mathcal{F}_t\}_{t\in[0,T]}$ is a right-continuous and $\p$-complete filtration that contains the available information up to time $t$. Throughout this paper, we use $\E[ \cdot ]$ to define the expectation under $\p$.
The number of individual claims occurring within the time horizon $[0,t]$ is modelled by the homogeneous Poisson process $\{N(t)\}_{t \in [0,T]}$ with intensity $\lambda>0$. For $i=1,\dots , n $, random variables $Y_i \in L_+^2$ denote the $i$th claim size, and $ Y_1 , Y_2 , \dots, $ are positive independent and identically distributed. Moreover, we assume that $\{N(t)\}_{t \in [0,T]}$ and $\{Y_i\}_{i=1,2,...}$ are stochastically independent, and then the compound Poisson process $\sum_{i=1}^{N(t)} Y_i$ is the size of the total claim that occurred up to time $t$.
The compound Poisson process is commonly used in the insurance literature to formulate the claim size. The filtration $\{\mathcal{F}_t\}_{t\in[0,T]}$ is generated by $\{N(t)\}_{t \in [0,T]}$ and $\{Y_i\}_{i=1,2,...}$ in the usual way.
Denote $ F(\cdot ) $ as the common cumulative distribution function (CDF) of $Y_i$, $i=1,2 , \dots $.
Assume $ a_Y := \E [Y]= \int_{\R^+} y \d F(y)<\infty$ and $\sigma_Y^2:=\int_{\R^+} y^2 \d F(y)<\infty$, that is, $ Y_i $ has finite first and second moments.

The occurrence time of the $i$-th claim is defined as $T_i=\inf\{t \le T, N(t)\ge i\}$. For $i=1,2,\dots $, given the occurrence time $T_i$ and the severity of the claim $Y_i$, the insurer will cede $l_1(T_i,Y_i)$ and $l_2(T_i,Y_i)$ to reinsurer 1 and reinsurer 2, respectively, where $l_j:[0,T]\times \R^+\to \R^+$ is the indemnity function of reinsurer $j$, $j=1,2$.
Consequently, the insurer retains the loss $Y_i-l_1(T_i,Y_i)-l_2(T_i,Y_i)$, $i=1,2,\dots$.
We use Poisson random measures to express the aggregate claims and denote the Poisson random measure by $N( \sd t, \d y)$ with the compensator $v(\d y)\d t :=\lambda \d F(y) \d t$.
Thus, the severity of the aggregate claim at time $t$ is $\sum^{N(t)}_{i=1} Y_i = \int^t_0 \int_{\R^+} y N(\sd s,\d y)$, and it has the expected value
$$\E \left [\sum^{N(t)}_{i=1}Y_i \right ] = \E\left[\int^t_0 \int_{\R^+} y N(\sd s,\d y)\right] = \int^t_0 \int_{\R^+} y v( \d y) \d s =\lambda a_Y t $$
Furthermore, the losses covered by reinsurer $j=1,2$ and the insurer can be written as
\begin{align*}
        \sum^{N(t)}_{i=1}  l_j \left (T_i,Y_i \right ) & = \int^t_0 \int_{\R^+} l_j(s,y) N( \sd s,\d y), ~~~ j =1 , 2 ,\\
      \sum^{N(t)}_{i=1} \left (Y_i-l_1(T_i,Y_i)-l_2(T_i,Y_i) \right ) & = \int^t_0 \int_{\R^+} \left  (y-l_1(s,y)-l_2(s,y) \right ) N( \sd s,\d y) .
\end{align*}

We suppose that the insurer adopts the expected value premium principle with constant safety loading $\theta>0$, and then the insurer will receive an instantaneous premium flow at the rate of $c:=(1+\theta)\lambda a_Y= (1+\theta) \int_{\R^+} y v( \sd y)$.
In contrast, the two reinsurers charge reinsurance premiums independently through different premium principles. 
We assume that reinsurer 1 adopts the mean-variance premium principle and receives instantaneous premium flow at the rate of
$$p_1(t)=\int_{\R^+} l_1(t,y)v( \sd y)+\xi_1(t) \int_{\R^+} l^2_1(t,y)v( \sd y),$$
where $\xi_1(t)$ is the security loading,
and reinsurer 2 adopts the expected-value premium principle with the instantaneous premium rate
$$p_2(t):=(1+\xi_2(t))\int_{\R^+} l_2(t,y) v( \sd y),$$
where $\xi_2(t)$ is the security loading.
In the present work, we assume that the reinsurance premium principles are fixed, while the reinsurers can adjust the reinsurance price by changing the security loadings. Thus, for $j=1,2$, the choices on $ \xi_j (t) $ reflect the attitude of the insurer $j$ toward the risk ceded from the insurer and the competition with the other reinsurer.
In the rest of the paper, we call $\xi_1( t)$ and $\xi_2( t)$ the \textit{reinsurance premium strategies}.
Furthermore, we assume that $\xi_1(t) \in [\frac{\theta a_Y}{\sigma^2_Y}, \frac{\eta a_Y}{\sigma^2_Y}]$ and  $\xi_2(t) \in [\theta, \eta]$, which ensure reinsurance price non-arbitrage and coverage that is not unduly expensive. 
After paying for the reinsurance contracts, the net premium rate received by the insurer becomes $c-p_1(t)-p_2(t)$.


Denote by $x_{0}^{I}>0$, $x_{0}^{R1}>0$ and $x_{0}^{R2}>0$
the initial assets of the insurer and two reinsurers, respectively.
The insurer and the reinsurers receive credit (resp.~debit) interest from their positive (resp.~negative) surpluses at interest rates $\rho_I(t)$, $\rho_{R1}(t)$ and $\rho_{R2}(t)$, respectively.
Assume that the three interest rates are positive, bounded and deterministic.
Thus, the surplus process for insurer  is
\begin{equation} \label{wealth1}
\begin{cases}
\sd X_I(t)=\left [c-p_1(t)-p_2(t)+\rho_I(t)X_I(t) \right ] \d t - \int_{\R^+} \left (y-l_1(t,y)-l_2(t,y) \right ) N( \sd t, \sd y), \\
X_I(0)=x_{0}^I.
\end{cases}
\end{equation}
The two reinsurers have surplus processes
\begin{equation} \label{wealth2}
\begin{cases}
\sd X_{R1}(t)= \left [p_1(t)+\rho_{R1}(t) X_{R1}(t) \right ] \d t - \int_{\R^+} l_1(t,y)N( \sd t, \sd y),\\
X_{R1}(0)=x_{0}^{R1},
\end{cases}
\end{equation}
and
\begin{equation}\label{wealth3}
\begin{cases}
 \sd X_{R2}(t)=\left [p_2(t)+\rho_{R2}(t) X_{R2}(t) \right ] \d t - \int_{\R^+} l_2(t,y)N( \sd t, \sd y), \\
 X_{R2}(0)=x_{0}^{R2}.
\end{cases}
\end{equation}

Throughout this paper, we refer to $\{l_1, l_2\} :=\{l_1(t,y), l_2(t,y)\}_{(t,y)\in[0,T]\times \mathbb{R^+}}$ as one reinsurance strategy for the insurer, and $\xi_j:={\xi_j (t)}_{t \in [0,T]}$ is a reinsurance premium strategy for the reinsurer $j$, $j=1,2$. In the following, we define admissible strategies.

\begin{definition}\label{admission}
A \textit{market strategy} $\{l_1,l_2,\xi_1,\xi_2\}$ is said to be \textit{admissible} if: \vspace{-0.1in}
\begin{enumerate}
\item [(i)]  
$l_1$ and $l_2$ are non-negative $\mathbb{F}$-predictable processes such that  $l_1(t,y) +l_2(t,y ) \in [0,y]$ for $(t,y) \in [0,T] \times \R^+$;  \vspace{-0.1in}
\item[(ii)]
$\xi_1$ and $\xi_2 $ are $\mathbb{F}$-predictable processes such that $\xi_1(t) \in  [\frac{\theta a_Y}{\sigma^2_Y}, \frac{\eta a_Y}{\sigma^2_Y}]$ and $\xi_2(t) \in  [\theta, \eta] $ for $t \in [0,T]$;  \vspace{-0.1in}
\item[(iii)]
associated with $\{l_1,l_2,\xi_1,\xi_2\}$, the surplus processes \eqref{wealth1}, \eqref{wealth2} and \eqref{wealth3} have unique strong solutions $X_I(\cdot)$, $X_{R1}(\cdot)$ and $X_{R2}(\cdot)$, respectively, which are $c \grave{a}dl \grave{a}g$, $\mathbb{F}$-adapted processes satisfying $\E[\sup_{t \in [0,T]} |X_k (t)|^2 ] < \infty$ for $k=I, R1, R2$.
\end{enumerate}
\end{definition}

Let $\mathscr{A}:=\mathscr{A}_I \times \mathscr{A}_{R1} \times \mathscr{A}_{R2}$ be the set including all admissible market strategies, where $ \mathscr{A}_j$ is the set of admissible strategies for participant $k=I,R1,R2$.
Note that $\{l_1,l_2,\xi_1,\xi_2\} \in \mathscr{A} $ if and only if $\{l_1, l_2 \} \in \mathscr{A}_I$, $\xi_1 \in \mathscr{A}_{R1}$ and $\xi_2 \in \mathscr{A}_{R2}$.

In this paper, we discuss this problem under a time-inconsistent mean-variance framework. Our aim is to find the optimal strategies to maximize the expectations and minimize the variance of the insurer's and reinsurers' respective terminal surpluses. We denote $\E_{t,x} [\cdot]:=\E[\cdot | \mathcal{F}_t, X(t)=x] $ and $\Var_{t,x} (\cdot):=\Var ( \cdot | \mathcal{F}_t, X(t)=x)$, where $X(t)=x$ stands for $(X_I(t),X_{R1}(t),X_{R2}(t))=(x_I,x_{R1},x_{R2})$. Let $\gamma_I > 0 $, $\gamma_{R1} > 0$ and $\gamma_{R2} > 0$, which reflect their attitudes towards the trade-off between risk and return. Then, we can obtain the objective function as follows:

\begin{definition}
Given $\{ l_1, l_2,\xi_1,\xi_2\} \in \mathscr{A}$,
the insurer's objective at time $t$ is
\begin{gather}
J_I(t, x_I; l_1, l_2,\xi_1,\xi_2)= \E_{t,x_I}[X_I(T)] -\frac{\gamma_I}{2} \Var_{t,x_I} ( X_I(T)) \label{ob1},
\end{gather}
reinsurer 1's objective at time $t$ is
\begin{gather}
J_{R1}(t, x_{R1}; l_1, \xi_1)= \E_{t,x_{R1}}[X_{R1}(T)] -\frac{\gamma_{R1}}{2} \Var_{t,x_{R1}} (X_{R1}(T)) \label{ob2},
\end{gather}
and reinsurer 2's objective at time $t$ is
\begin{gather}
J_{R2}(t, x_{R2}; l_2,\xi_2)= \E_{t,x_{R2}}[X_{R2}(T)] -\frac{\gamma_{R2}}{2} \Var_{t,x_{R2}} (X_{R2}(T))\label{ob3}.
\end{gather}
\end{definition}

Now, we use the Stackelberg game to model the relationship between the insurer and reinsurers.
Being aware of the insurer's response, reinsurers play the leadership role and determine the reinsurance premium first. \textcolor{r}{In contrast, the insurer as a price taker, would be the follower in the game. And it has the right to choose a reinsurance plan under the given reinsurance premium.
The competition between two reinsurers is described by the Nash game}. Reinsurer 1 and reinsurer 2 are two oligarchs in the reinsurance market. They compete with each other and simultaneously determine their reinsurance premiums.
Specifically, given reinsurance premium strategies $\xi_1 \in \mathscr{A}_{R1}$ and $\xi_2\in \mathscr{A}_{R2}$ for reinsurer 1 and reinsurer 2, the insurer can choose an optimal reinsurance strategy $\{l_1^* (\cdot;\xi_1,\xi_2) , l_2^*(\cdot;\xi_1,\xi_2) \} \in \mathscr{A}_I$ based upon her objective. In anticipation of the insurer's choice of the optimal reinsurance strategy, two reinsurers simultaneously determine their reinsurance premium strategies $\xi_1^*$ and $\xi_2^*$.
\textcolor{r}{The search process of the equilibrium reinsurance and reinsurance price on the reinsurance market can be described by the following three steps.}
\begin{enumerate}
\item[1.]
In the first step, we search for an admissible reinsurance strategy to optimize the insurer's objective function for any reinsurer strategy set $ \{\xi_1, \xi_2 \} \in \mathscr{A}_{R1} \times \mathscr{A}_{R2}$.
The insurer's optimal strategy is associated with the given reinsurance premium strategies $\xi_1$ and $\xi_2$ and is denoted by $\{l_1^* (\cdot;\xi_1,\xi_2) , l_2^*(\cdot;\xi_1,\xi_2) \} $.
The insurer's response $\{l_1^* (\cdot;\xi_1,\xi_2) , l_2^*(\cdot;\xi_1,\xi_2) \} $ is available to both reinsurers and will be used by the reinsurers to design their own strategies in the following steps.
\vspace{-0.1in}
\item[2.]
In the second step, we find the reinsurance premium strategies $\xi_1^*(\cdot;\xi_2)$ and $\xi_2^*(\cdot;\xi_1)$ for the two reinsurers when the other reinsurer's strategy is given and the insurer's choice is the optimal reinsurance strategy we found in the first step. \vspace{-0.1in}
\item[3.]
\textcolor{r}{In the third step, we find the equilibrium reinsurance premium strategies for the Nash game between two reinsurers,} which is $\{\xi^*_1,\xi^*_2\}$ from the reaction function we obtain in the second step. Then, we substitute the equilibrium reinsurance strategy into $\{l_1^* (\cdot;\xi_1,\xi_2) , l_2^*(\cdot;\xi_1,\xi_2 ) \}$ to obtain the equilibrium solution for the Stackelberg game $\{l_1^*(\cdot),l_2^*(\cdot)\}$.
\end{enumerate}

%

\section{Time-consistent strategy}\label{time-consitent}
\subsection{Equilibrium strategy}

Since the dynamic mean-variance problem of \eqref{ob1}-\eqref{ob3} is essentially time inconsistent, we cannot directly use Bellman's optimality principle. To overcome this difficulty, we address this problem within a game-theoretic framework and search for the Nash subgame perfect equilibrium strategy, which was pioneered by \cite{RMyopia}. We can derive the extended Hamilton-Jacobi-Bellman equation following the method introduced by \cite{tomas2017}.

\begin{definition}\label{def:equilibrium stategy}
Given the objective function $J_I$, $J_{R1}$ and $J_{R2}$, the equilibrium reinsurance strategy and equilibrium reinsurance premium strategy are defined as following.
\begin{enumerate}[(i)]
\item Given $\xi_1 \in \mathscr{A}_{R1}$ and $\xi_2 \in \mathscr{A}_{R2}$, let $ \{l_1^* (\cdot;\xi_1,\xi_2) , l_2^*(\cdot;\xi_1,\xi_2) \} \in \mathscr{A}_I$
be a reinsurance strategy associated with a given $ (\xi_1 , \xi_2 ) $. 
For any $(t,y) \in [0,T] \times \R^+$ and fixed real number $\epsilon>0$, define the associated reinsurance strategy as:
\begin{equation*}
l_j^{\epsilon}(s,y;\xi_1,\xi_2)=\begin{cases}
\tilde{l}_j(y) , & s \in \left[t,t+\epsilon\right) \\
l_j^* (s, y ;\xi_1,\xi_2) , & s \in \left[ t+\epsilon,T \right)
\end{cases} , ~~~~~~~~j=1,2 ,
\end{equation*}
where $\tilde{l}_1: \R^+ \rightarrow \R^+ $ and $\tilde{l}_2: \R^+ \rightarrow \R^+ $ are functions such that $\{l_1^{\epsilon}(\cdot;\xi_1,\xi_2),l_2^{\epsilon}(\cdot;\xi_1,\xi_2)\} \in \mathscr{A}_I$.
If
\begin{gather*}
\lim_{\epsilon \to 0^+} \inf \frac{1}{\epsilon}[J_I(t, x_I; l_1^*(\cdot; \xi_1,\xi_2), l_2^*(\cdot; \xi_1,\xi_2),\xi_1,\xi_2)-J_I(t, x_I; l_1^{\epsilon}(\cdot; \xi_1,\xi_2), l_2^{\epsilon}(\cdot; \xi_1,\xi_2),\xi_1,\xi_2)]>0,
\end{gather*}
then, $\{l_1^*(\cdot;\xi_1,\xi_2) , l_2^*(\cdot;\xi_1,\xi_2)\}$ is called the equilibrium reinsurance strategy associated with $\xi_1$ and $\xi_2$, and $V_I(t,x_I;\xi_1,\xi_2)=J_I(t, x_I; l_1^*(\cdot;\xi_1,\xi_2) , l_2^*(\cdot;\xi_1,\xi_2),\xi_1,\xi_2)$ is the equilibrium value function for insurer associated with $ (\xi_1 , \xi_2 ) $.

\item Given $\xi_2 \in \mathscr{A}_2 $, let $ \xi_1^*(\cdot;\xi_2) \in \mathscr{A}_{R1}$
be a reinsurance premium strategy for reinsurer 1 associated with a given $\xi_2 $. For any $t \in [0,T] $ and fixed real number $\epsilon>0$, define the associated reinsurance premium strategy as:
\begin{equation*}
\xi_1^{\epsilon}(s;\xi_2)=\begin{cases}
\tilde{\xi}_1 , & s \in \left[t,t+\epsilon\right) \\
\xi_1^* (s;\xi_2) , & s \in \left[ t+\epsilon,T \right)
\end{cases}
\end{equation*}
where $\tilde{\xi}_1 $ is a real number such that $\xi_1^{\epsilon}(\cdot;\xi_2) \in \mathscr{A}_{R1}$. If
\begin{gather*}
\lim_{\epsilon \to 0^+} \inf \frac{1}{\epsilon} \left [J_{R1}(t, x_{R1}; l_1^*(\cdot; \xi_1^*(\cdot;\xi_2 ),\xi_2 ), \xi_1^*(\cdot;\xi_2 ))
-J_{R1}(t, x_{R1}; l_1^*(\cdot; \xi_1^{\epsilon}(\cdot;\xi_2 ), \xi_1^{\epsilon}(\cdot;\xi_2 ) )\right  ]  \ge 0,
\end{gather*}
then we call $\xi_1^*(\cdot;\xi_2)$ the equilibrium reinsurance premium strategy for reinsurer 1 associated with $\xi_2$ and $V_{R1}(t,x_{R1};\xi_2)=J_{R1}(t, x_{R1};l_1^*(\cdot;\xi_1^*(\cdot;\xi_2),\xi_2) , \xi_1^*(\cdot;\xi_2))$ the equilibrium value function for reinsurer 1 associated with $\xi_2$.

\item Given $\xi_1 \in \mathscr{A}_{R1}$, let $ \xi_2^*(\cdot;\xi_1) \in \mathscr{A}_{R2}$
be the reinsurance premium strategy for reinsurer 2 associated with a given $\xi_1 $. For any $t \in [0,T] $ and fixed real number $\epsilon>0$, define the associated reinsurance premium strategy as:
\begin{equation*}
\xi_2^{\epsilon}(s;\xi_1)=\begin{cases}
\tilde{\xi}_2 , & s \in \left[t,t+\epsilon\right) \\
\xi_2^* (s, ;\xi_1) , & s \in \left[ t+\epsilon,T \right)
\end{cases}
\end{equation*}
where $\tilde{\xi}_2$ is a real number such that $\xi_2^{\epsilon}(\cdot;\xi_1) \in \mathscr{A}_{R2}$. If
\begin{gather*}
\lim_{\epsilon \to 0^+} \inf \frac{1}{\epsilon} \left [J_{R2}(t, x_{R2}; l_2^*(\cdot; \xi_1 ,\xi_2^*(\cdot;\xi_1 )),\xi_2^*(\cdot;\xi_1 ))
-J_{R2}(t, x_{R2};
l_2^*(\cdot; \xi_1 ,\xi_2^{\epsilon}(\cdot;\xi_1 ) ,\xi_2^{\epsilon}(\cdot;\xi_1 ))\right ]  \ge 0,
\end{gather*}
then we call $\xi_2^*(\cdot;\xi_1)$ the equilibrium reinsurance premium strategy for reinsurer 2 associated with $\xi_1$ and $V_{R2}(t,x_{R2};\xi_1)=J_{R2}(t, x_{R2}; l_2^*(\cdot;\xi_1,\xi_2^*(\cdot;\xi_1)),\xi_2^*(\cdot;\xi_1))$ the equilibrium value function for reinsurer 2 associated with $\xi_1$.

\item \textcolor{r}{$( \xi_1^*(\cdot),\xi_2^*(\cdot) )$ are called the equilibrium reinsurance premium strategies if $\xi_1^*(\cdot)$ is the equilibrium reinsurance premium strategy for reinsurer 1 associated with $\xi_2^*(\cdot)$ and $\xi_2^*(\cdot)$ is the equilibrium reinsurance premium strategy for reinsurer 1 associated with $\xi_1^*(\cdot)$. More specifically, } $( \xi_1^*(\cdot),\xi_2^*(\cdot) )$ are called the equilibrium reinsurance premium strategies if
\begin{equation} \label{equilibriumxi}
\begin{cases}
\xi_1^*(t)=\xi_1^*(t; \xi_2^*(t)),\\
\xi_2^*(t)=\xi_2^*(t; \xi_1^*(t)),
\end{cases}~~~~\text{ for all } t \in [0,T].
\end{equation}
Then, the equilibrium reinsurance strategy for the insurer is $ ( l_1^*,l_2^* ) = ( l_1^*(\cdot;\xi_1^*,\xi_2^*),l_2^*(\cdot;\xi_1^*,\xi_2^*) ) $.
Moreover, $V_I(t,x_I)=V_I(t,x_I;\xi_1^*,\xi_2^*)$, $V_{R1}(t,x_{R1})=V_{R1}(t,x_{R1};\xi_2^*)$ and $V_{R2}(t,x_{R2})=V_{R2}(t,x_{R2};\xi_1^*)$ are the equilibrium value functions for insurer, reinsurer 1 and reinsurer 2, respectively.
\end{enumerate}
\end{definition}

The value functions of insurer (I), reinsurer 1 (R1) and reinsurer 2 (R2) can also be written as
$V_I(t, x_I)=J_I(t, x_I; l^*_1, l^*_2,\xi^*_1,\xi^*_2)$, $V_{R1}(t, x_{R1})=J_{R1}(t, x_{R1}; l^*_1,\xi^*_1)$ and $V_{R2}(t, x_{R2})=J_{R2}(t, x_{R2};  l^*_2,\xi^*_2)$.

\begin{proposition} \label{pro: independent}
The objective functions of the insurer and reinsurers (\ref{ob1})-(\ref{ob3}) are separable in their own surpluses and independent of the opponent's. Moreover, neither the insurer's nor reinsurers' optimal strategies depend on the surpluses.
\end{proposition}

\newcommand\deref[1]{definition~\ref{#1}}
\newcommand\prref[1]{proposition~\ref{#1}}

The proof of \prref{pro: independent} is in appendix \ref{proof-independ}. However, to use the equilibrium strategy defined in \deref{def:equilibrium stategy} and derive the extended HJB equation for the value function, we need to demonstrate that the optimal strategies of the insurer and reinsurers are state independent, which is true according to \prref{pro: independent}. Intuitively, it is difficult for one participant to acquire and make decisions based on the information on the surplus levels of other participants.
Therefore, we can concentrate on equilibrium strategies given in \deref{def:equilibrium stategy} and apply the extended HJB equation to solve the mean-variance problem \eqref{ob1}-\eqref{ob3}.

\subsection{The extended HJB equation}

First, for an arbitrary function $\varphi: [0,T] \times \R  \to  \R$ with partial derivatives $ \frac{\partial \varphi(t,x)}{\partial t} $ and $ \frac{\partial \varphi(t,x)}{\partial x} $, we define the infinitesimal generators for the insurer's and two reinsurers' optimal problems acting on $\varphi$ as 

\begin{equation} \label{gen1}
\begin{split}
\mathscr{L}_I^{l_1,l_2,\xi_1,\xi_2}[\varphi(t,x_I)] = & \, \frac{\partial \varphi(t,x_I)}{\partial t} +
 \frac{\partial \varphi(t,x_I)}{\partial x} [ c-p_1(t)-p_2(t)+\rho_I(t)x_I]\\
 & \, +\int_{\R^+}  \left [ \varphi ( t,x_I-(y-l_1(t,y)-l_2(t,y)))-\varphi(t,x_I) \right ] v(\sd y) ,
 \end{split}
 \end{equation}

\begin{equation}\label{gen2}
 \begin{split}
\mathscr{L}_{R1}^{l_1,\xi_1}[\varphi(t,x_{R1})] = &\,
\frac{\partial \varphi(t,x_{R1})}{\partial t} +
 \frac{\partial \varphi(t,x_{R1})}{\partial x} \Bigl[ \int_{\R^+} l_1(t,y) v(\sd y)  +
 \xi_1(t) \int_{\R^+} l_1^2(t,y) v(\sd y) \\
 & \,+ \rho _{R1}(t) x_{R1} \Bigr]+\int_{\R^+} \left [ \varphi ( t,l_1(t,y) ) -\varphi(t,x_{R1})  \right ] v(\sd y),
 \end{split}
\end{equation}

\begin{equation} \label{gen3}
\begin{split}
\mathscr{L}_{R2}^{l_2,\xi_2}[\varphi(t,x_{R2})] = & \,
\frac{\partial \varphi(t,x_{R2})}{\partial t} +
 \frac{\partial \varphi(t,x_{R2})}{\partial x} \Bigl[ (1+\xi_{2}(t))\int_{\R^+} l_2(t,y) v(\sd y)  + \rho _{R2}(t) x_{R2} \Bigr]\\
& \, +\int_{\R^+} \left [ \varphi ( t,x
_{R2}-l_2(t,y)) -\varphi(t,x_{R2}) \right ] v(\sd y).
 \end{split}
\end{equation}

\begin{theorem}[Verification Theorem]~
\begin{itemize}\item[(i)]
Given $\{\xi_1,\xi_2\} \in \mathscr{A}_{R1} \times \mathscr{A}_{R2}$, if there exist real value function $V_I(t,x_I; \xi_1,\xi_2) \in C^{1,2}([0,T]\times \R^+)$ and \textcolor{r}{real function} $g(t,x_I;\xi_1,\xi_2) \in C^{1,2}([0,T]\times \R^+)$ satisfying the following conditions:
\begin{equation} \label{sup1}
\begin{cases}
\sup\limits_{\{l_1,l_2\} \in \mathscr{A}_I } \Bigl\{\mathrm{HJB}_I ( t, x_I ;l_1, l_2 , \xi_1, \xi_2 )\Bigr\}=0,\\
 \mathscr{L}_I^{l^*_1,l^*_2,\xi_1,\xi_2}[g_I(t,x_I; \xi_1 ,\xi_2 )]=0,\\
 V_I(T,x_I;\xi_1 ,\xi_2 )=x_I,\\
 g_I(T,x_I; \xi_1 ,\xi_2 )=x_I,
\end{cases}
\end{equation}
where
\begin{align*}
 \mathrm{HJB}_I (t, x_I ;l_1, l_2 , \xi_1, \xi_2  )  : =   & \, \mathscr{L}_I^{l_1,l_2,\xi_1,\xi_2}[V_I(t,x_I; \xi_1,\xi_2)]- \frac{\gamma_I}{2} \mathscr{L}_I^{l_1,l_2,\xi_1,\xi_2}[g_I^2(t, x_I; \xi_1 ,\xi_2 )] \\
 &\, +\gamma_I g_I(t,x_I; \xi_1 ,\xi_2 ) \mathscr{L}_I^{l_1,l_2,\xi_1,\xi_2}[g_I(t,x_I;\xi_1 ,\xi_2 )] \\
\{l^*_1(t,y; \xi_1,\xi_2), l^*_2(t,y;\xi_1,\xi_2)\} : = & \, \arg\sup\limits_{\{l_1,l_2\} \in \mathscr{A}_I }  \mathrm{HJB}_I (t, x_I ;l_1, l_2 , \xi_1, \xi_2 ),
\end{align*}
then $V_I(t, x_I;\xi_1, \xi_2)=J_I(t, x_I; l_1^*(\cdot;\xi_1,\xi_2)$, which is the equilibrium value function for insurer associated with $\xi_1$ and $\xi_2$, and $\{l^*_1(\cdot;\xi_1,\xi_2), l^*_2(\cdot;\xi_1,\xi_2)\}$ is the equilibrium reinsurance strategy associated with $\xi_1$ and $\xi_2$.

\item[(ii)]
Given $\xi_2 \in \mathscr{A}_{R2}$, if there exist real value function $V_{R1}(t,x_{R1};\xi_2) \in C^{1,2}([0,T]\times \R^+)$ and \textcolor{r}{real function} $g_{R1}(t,x_{R1};\xi_2) \in C^{1,2}([0,T]\times \R^+)$ satisfying the following conditions:
\begin{equation} \label{sup2}
\begin{cases}
\sup\limits_{\xi_1 \in \mathscr{A}_{R1} } \Bigl\{ \mathrm{HJB}_{R1} (t , x_{R1} ; l_1^*(\cdot:\xi_1,\xi_2), \xi_1  )   \Bigr\}=0, \\
\mathscr{L}_{R1}^{l^*_1(\cdot;\xi_1^*,\xi_2),\xi^*_1}[g_{R1}(t,x_{R1};\xi_2)]=0,\\
 V_{R1}(T,x_{R1};\xi_2)=x_{R1},\\
 g_{R1}(T,x_{R1};\xi_2)=x_{R1},
\end{cases}
\end{equation}
where
\begin{align*}
 \mathrm{HJB}_{R1} (t , x_{R1} ; l_1^*(\cdot:\xi_1,\xi_2), \xi_1 ) : =&\, \mathscr{L}_{R1}^{ l_1^*(\cdot:\xi_1,\xi_2),\xi_1}[V_{R1}(t,x_{R1};\xi_2)]- \frac{\gamma_{R1}}{2} \mathscr{L}_{R1}^{ l_1^*(\cdot:\xi_1,\xi_2),\xi_1}[g_{R1}^2(t, x_{R1};\xi_2)] \\
& \, +\gamma_{R1} g_{R1}(t,x_{R1};\xi_2) \mathscr{L}_{R1}^{ l_1^*(\cdot:\xi_1,\xi_2),\xi_1}[g_{R1}(t,x_{R1};\xi_2)]\\
\xi^*_1(\cdot;\xi_2) ,
: = & \, \arg\sup\limits_{\xi_1 \in \mathscr{A}_{R1} } \Bigl\{  \mathrm{HJB}_{R1} (t , x_{R1} ;  l_1^*(\cdot:\xi_1,\xi_2) , \xi_1 ) \Bigr\},
 \end{align*}
then $V_{R1}(t, x_{R1};\xi_2)=J_{R_1}(t,x_{R1};l^*_1(\cdot;\xi_1^*(\cdot;\xi_2),\xi_2),\xi_1^*(\cdot;\xi_2))$, which is the equilibrium value function for reinsurer 1 associated with $\xi_2$, and $\xi^*_1(\cdot;\xi_2)$ is the equilibrium reinsurance premium strategy for reinsurer 1 associated with $\xi_2$.
\item[(iii)]
Given $\xi_1 \in \mathscr{A}_{R1}$, if there exist real value function $V_{R2}(t,x_{R2};\xi_1) \in C^{1,2}([0,T]\times \R^+)$ and \textcolor{r}{real function} $g_{R2}(t,x_{R2};\xi_1) \in C^{1,2}([0,T]\times \R^+)$ satisfying the following conditions:
\begin{equation} \label{sup3}
\begin{cases}
\sup\limits_{\xi_1 \in \mathscr{A}_{R2} } \Bigl\{  \mathrm{HJB}_{R2} (t , x_{R2} ;  l_2 ^*(\cdot:\xi_1,\xi_2) , \xi_2 )  \Bigr\}=0\\
 \mathscr{L}_{R2}^{ l_2 ^*(\cdot:\xi_1^*,\xi_2) , \xi_2 }[g_{R2}(t,x_{R2};\xi_1)]=0,\\
 V_{R2}(T,x_{R2};\xi_1)=x_{R2},\\
 g_{R2}(T,x_{R2};\xi_1)=x_{R2},
\end{cases}
\end{equation}
where
\begin{align*}
 \mathrm{HJB}_{R2} (t , x_{R2} ;  l_2 ^*(\cdot:\xi_1,\xi_2) , \xi_2  )  : = &\,  \mathscr{L}_{R2}^{ l_2 ^*(\cdot:\xi_1,\xi_2) , \xi_2 }[V_{R2}(t,x_{R2};\xi_1)]- \frac{\gamma_{R2}}{2} \mathscr{L}_{R2}^{ l_2 ^*(\cdot:\xi_1,\xi_2) , \xi_2 }[g_{R2}^2(t, x_{R2};\xi_1)] \\
&\, +\gamma_{R2} g_{R2}(t,x_{R2};\xi_1) \mathscr{L}_{R2}^{ l_2 ^*(\cdot:\xi_1,\xi_2) , \xi_2 }[g_{R2}(t,x_{R2};\xi_1)]\\
\xi^*_2( \cdot ;\xi_1)  : =& \, \arg\sup\limits_{\xi_1 \in \mathscr{A}_{R2} } \Bigl\{ \mathrm{HJB}_{R2} (t , x_{R2} ; l_2 ^*(\cdot:\xi_1,\xi_2) , \xi_2  ) \Bigr\},
\end{align*}
then $V_{R2}(t, x_{R2};\xi_1)=J_{R_2}(t,x_{R2};l^*_2(\cdot;\xi_1,\xi_2^*(\cdot;\xi_1)),\xi_2^*(\cdot;\xi_1))$, which is the equilibrium value function for reinsurer 2 associated with $\xi_1$, and $\xi^*_2(\cdot;\xi_1)$ is the equilibrium reinsurance premium strategy for reinsurer 1 associated with $\xi_1$.
\end{itemize}
\end{theorem}

The proof of the verification theorem can be adapted from Theorem 5.2 of \cite{tomas2017} and Theorem 4.1 of \cite{bjork2010mu}. Here, we omit the proof.

If there exist $ ( \xi^*_1 ,\xi^*_2 ) \in \mathscr{A}_{R1} \times \mathscr{A}_{R2}$ such that $ \xi_1^*(t) = \xi^*_1(t, \xi^*_2(t))$ and $ \xi_2^* (t) = \xi^*_2(t,\xi^*_1(t))$ for all $ t\in [0,T]$, i.e., equation \eqref{equilibriumxi} is satisfied, then
$ ( \xi^*_1 ,\xi^*_2 )$ is the equilibrium reinsurance premium strategy. The value functions for the insurer and two reinsurers are $V_I(t,x_{I})=V_{I}(t,x_{I};\xi_1^*,\xi_2^*)$, $V_{R1}(t, x_{R1})=V_{R1}(t, x_{R1};\xi^*_2)$ and  $V_{R2}(t, x_{R2})=V_{R2}(t, x_{R2};\xi^*_1)$, respectively.

\subsection{Solution}

\begin{proposition} \label{inpro}
Given $( \xi_1 ,\xi_2 ) \in \mathscr{A}_{R1} \times \mathscr{A}_{R2}$, the associated equilibrium reinsurance strategies are
\begin{align*}
l^*_1(t, y;\xi_1,\xi_2)&=
\begin{cases}
q(t)y,& y\leq d(t),\\
\frac{\xi_2(t)}{2\xi_1(t)},& y>d(t),
\end{cases}
~~\text{ and }~~~
l^*_2(t, y;\xi_1,\xi_2) =
\begin{cases}
0,& y\leq d(t),\\
y-d(t),& y>d(t),
\end{cases}
\end{align*}
where
\begin{align}\label{def:q-and-d}
q(t)=\frac{\gamma_I e^{\int_t^T \rho_I(s) \sd s}}{2\xi_1(t)+\gamma_I e^{\int_t^T \rho_I(s) \sd s }} ~~ \text{ and } ~~~d(t)=\frac{\xi_2(t)}{\gamma_I e^{\int_t^T \rho(s)\sd s}}+\frac{\xi_2(t)}{2\xi_1(t)} .
\end{align}
\end{proposition}

The proof is given in Appendix \ref{proof-inpro}. \textcolor{r}{And the proof has promised that $\{l_1^*,l_2^*\}$ is admissible.}

The results in Proposition \ref{inpro} can be interpreted in the following way. At time $t$, the insurer holds a proportion reinsurance offered by reinsurer 1 and a stop-loss reinsurance offered by reinsurer 2. In the proportion reinsurance, the ceding proportion is $q(t)$ and the limit is $\frac{\xi_2(t)}{2\xi_1(t)}$. The stop-loss reinsurance is characterized by the deductible $d(t)$.
Suppose that the realized claim amount for the insurer at time $t$ is $y$. If $y $ is no larger than $d(t)$, the insurer retains the amount of $(1-q(t))y$ and transfers the amount of $q(t)y$ to reinsurer 1. If $y$ is strictly larger than $ d(t) $, the insurer transfers the amount of $\frac{\xi_2(t)}{2\xi_1(t)}$ to reinsurer 1 and $y-d(t)$ to reinsurer 2 and thus retains the amount of $\frac{\xi_2(t)}{\gamma_I e^{\int_t^T \rho(s)\sd s}}$.


Given $(\xi_1, \xi_2 )$, suppose that the associated equilibrium value functions of two reinsurers are
\begin{align*}
          V_{R1}(t,x_{R1};\xi_2)=e^{\int_t^T \rho_{R1}(s)\sd s} x_{R1} + B_{R1}(t) ~~~\text{ and }~~ V_{R2}(t,x_{R2};\xi_1)=e^{\int_t^T \rho_{R2}(s)\sd s} x_{R2} + B_{R2}(t) .
\end{align*}
Moreover, we assume
\begin{align*}
           g_{R1}(t,x_{R1};\xi_2) & =e^{\int_t^T \rho_{R1}(s)\sd s} x_{R1} + b_{R1}(t) ~~~\text{ and }~~
           g_{R2}(t,x_{R2};\xi_1)  =e^{\int_t^T \rho_{R2}(s)\sd s} x_{R2} + b_{R2}(t).
\end{align*}
Then, the extended HJB equation \eqref{sup2}-\eqref{sup3} can be written as
\begin{equation}\label{re2}
\sup_{\xi_1 \in \mathscr{A}_{R1}}\bigg\{B_{R1}'(t)
+e^{\int_t^T\rho_{R1}(s)\sd s}\Big[\int_{\R^+}[\xi_1(t)-\frac{\gamma_{R1}}{2}e^{\int_t^T\rho_{R1}(s)\sd s}]l_1^{*2}(t,y;\xi_1,\xi_2)v(\d y)\Big]\bigg\}, \end{equation}
and
\begin{equation}\label{re3}
\sup_{\xi_2 \in \mathscr{A}_{R2}}\bigg\{B_{R2}'(t)+e^{\int_t^T\rho_{R2}(s)\sd s}\Big[\int_{\R^+}\xi_2(t)l_2^*(t,y;\xi_1,\xi_2)-\frac{\gamma_{R2}}{2}e^{\int_t^T\rho_{R2}(s)\sd s}l^{*2}_2(t,y;\xi_1,\xi_2)v(\d y)\Big]\bigg\}.
\end{equation}
The equilibrium reinsurance premium strategies for two reinsurers, given their competitor's premium strategy, can be calculated from the first-order conditions of \eqref{re2} and \eqref{re3}, which are
\begin{align}\label{firstorder1}
\Lambda_{R1}(\xi_1,\xi_2) & =  \left [2q(t)\left (\frac{\gamma_{R1}e^{\int_t^T \rho_{R1}(s) \sd s}}{\gamma_I e^{\int_t^T \rho_I(s) \sd s}}+1\right )-1\right ]\int_0^{d(t)} y^2 \d F(y)  + d(t)^2\left [ \frac{\gamma_{R1}e^{\int_t^T \rho_{R1}(s) \sd s}}{\xi_1}-1\right  ]S( d(t)) \notag  \\
&  =0
\end{align}
and
\begin{align} \label{firstorder2}
\Lambda_{R2}(\xi_1,\xi_2) & =  \left [1+\frac{\gamma_{R2}e^{\int_t^T \rho_{R2}(s) \sd s}}{\gamma_I e^{\int_t^T \rho_I(s) \sd s}}+\frac{\gamma_{R2}e^{\int_t^T \rho_{R2}(s) \sd s}}{2\xi_1}\right ] \int_{d(t)}^{+\infty} (y-d(t)) \d F(y) -d(t) S (d(t)) \notag \\
& = 0 ,
\end{align}
where $  S (d(t)) =1 - F(d(t)) $ and functions $ q(t) $ and $ d(t) $ are defined in \eqref{def:q-and-d}.
%
%
%
We can find $\xi_1^*(\cdot;\xi_2)$ from \eqref{firstorder1} and $\xi_2^*(\cdot;\xi_1)$ from \eqref{firstorder2}. If we can find a pair of $(\xi_1^*,\xi_2^*)$ satisfying both \eqref{firstorder1} and \eqref{firstorder2}, they are indeed the equilibrium reinsurance premium at time $t$. From the equilibrium reinsurance premium strategy $\{\xi_1^*(t),\xi_2^*(t)\}$, we can further obtain the equilibrium reinsurance strategy $\{l_1^*(\cdot),l_2^*(\cdot)\}$ for the insurer by substituting $\{\xi_1^*(t),\xi_2^*(t)\}$ into $\{l^*_1(t, y;\xi_1,\xi_2),l^*_2(t, y;\xi_1,\xi_2)\}$.

\section{Example: optimal reinsurance strategies under exponential claims}\label{example}
In section \ref{time-consitent}, we obtain the explicit form of $\{l^*_1(t, y;\xi_1,\xi_2),l^*_2(t, y;\xi_1,\xi_2)\}$. However, without information on the distribution of $Y_i$, we cannot solve \eqref{firstorder1} and \eqref{firstorder2} to obtain an explicit form of $\{\xi_1^*(\cdot;\xi_2),\xi_2^*(\cdot;\xi_1)\}$. 
To better understand the reinsurance premium strategies in a market equilibrium, in this section, we impose the assumption that the claim sizes $Y_i$, $i=1,2,...$, follow the exponential distribution with expectation $ 1 / \beta $, i.e., $Y_i \sim \exp(\beta)$.\footnote{$Y_i $ has density function  $f(y)=\beta e^{-\beta y}$, distribution function $F(y)=1-e^{-\beta y}$, expectation $ 1 /\beta $, and variance $ 1 /\beta^2 $. }
Then, we explicitly determine $\{\xi_1^*(\cdot),\xi_2^*(\cdot)\}$.


In the first attempt, we assume that $\xi_1^*(\cdot)$ and $\xi_2^*(\cdot)$ can take any positive values, i.e., $ \xi_i^* : [0,T] \rightarrow [0, \infty)$, $i=1,2$.
Fix $ t \in [0,T]$; we can determine the reaction functions $\xi_1( \cdot; \xi_2)$ and $\xi_2(\cdot; \xi_1)$ from \eqref{firstorder1} and \eqref{firstorder2} for two reinsurers given the strategy of their competitor.
Consequently, $\xi_1^*(t)$ and $\xi_2^*(t)$ are the coordinates of the intersection point of the two reaction functions. 
Since $\xi_1^*(t)$ and $\xi_2^*(t)$ can be determined in the same way for all $ t \in [0,T]$, in the following discussion, we focus on a given time $t $. For notational simplicity,
we denote $d=d(\xi_1,\xi_2)>0$, 
\begin{align*}
          C_{j} & =\frac{\gamma_{j}e^{\int_t^T\rho_{j}(s) \d s}}{\gamma_{I}e^{\int_t^T\rho_{I}(s) \d s}}, ~~~j = R1 , R2 \\
          B_{R1} &=\frac{\gamma_{R1}e^{\int_t^T\rho_{R1}(s) \d s}}{\xi_1} ~~~\text{ and }~~~B_{R2}=\frac{\gamma_{R2}e^{\int_t^T\rho_{R2}(s) \d s}}{\xi_2} .
\end{align*}
As we have $f(y)=\beta e^{-\beta y}$, we can rewrite \eqref{firstorder1} as the following quadratic equation of $B_{R1}$
\begin{align*}
\left [\frac{B_{R1}(2C_{R1}+1)-2C_{R1}}{2C_{R1}+B_{R1}}\right] e(\beta d)+B_{R1}-1=0, ~~\text{ where }  e(\beta d) : = \left [\frac{2}{(\beta d)^2}e^{\beta d}- \left (1+\frac{2}{\beta d}+\frac{2}{(\beta d)^2} \right ) \right ] .
\end{align*}


Since $-2C_{R1}(e(\beta d)+1)<0$, the above equation has one positive and one negative solution. In particular, the positive solution is $B_{R1}=G( \beta d  )$, where $ G(x) = g (e(x)) $ and
\begin{align*}
        g(x) : = \frac{1}{2} \left ( -[2 C_{R1}-1+x(2C_{R1}+1)]+\sqrt{[2C_{R1}-1+x(2C_{R1}+1)]^2+8C_{R1}(x+1)} \right ) .
\end{align*}
%
%
%
Now, we establish a relation between $\xi_1$ and $\xi_2$ from \eqref{firstorder1} via the following equation:
\begin{equation}\label{xi1-2-1}
           B_{R1} = \frac{\gamma_{R1}e^{\int_t^T\rho_{R1}(s) \d s}}{\xi_1} = G(\beta d ) =G \left ( \frac{\beta \xi_2}{\gamma_Ie^{\int_t^T\rho_I(s) \d s}}+\frac{\beta \xi_2}{2\xi_1} \right ) .
\end{equation}

\begin{lemma} \label{ghG}
The function $e(x)$ is continuously increasing in $x \in ( 0,+\infty)$, $g(x)$ is continuously decreasing in $ x\in (0,+\infty)$, and $G(x)$ is continuously decreasing in $ x\in (0,+\infty)$.
\end{lemma}

Since $G(x)$ is a decreasing function, its inverse function $ G^{-1}$ is well defined. Now, we represent $\xi_2$ as a function of $\xi_1$, say $ h_1 (\xi_1 )$, from \eqref{xi1-2-1} as
\[\xi_2 = h_1(\xi_1) := \frac{\gamma_I e^{\int_t^T \rho_I(s) \d s}G^{-1}(\frac{\gamma_{R1} e^{\int_t^T \rho_{R1}(s) \d s} }{\xi_1} )}{\beta \left (\frac{\gamma_I e^{\int_t^T \rho_I(s) \d s}}{2\xi_1}+1 \right )} . \]
It is clear that $h_1(\xi_1)$ is decreasing in $\xi_1$ because $G^{-1}(\cdot)$ is a decreasing function.

Similarly, under the exponential distribution, \eqref{firstorder2} implies
\begin{equation}\label{xi1-2-2}
\xi_2 = h_2 (\xi_1 ) : =\frac{\gamma_I e^{\int_t^T \rho_I(s) \d s}}{\beta}\left ( 1+C_{R2} -\frac{\gamma_I e^{\int_t^T \rho_I(s) \d s}}{\gamma_I e^{\int_t^T \rho(s) \d s}+2\xi_1} \right )
\end{equation}
and $h_2(\xi_1)$ is increasing in $\xi_1$.

\begin{proposition} \label{exist}
Assume that $Y_i \sim \exp(\beta) $, $i=1,2,...$, and $ \xi_j (t) \geq 0 $ for $ j=1,2$ and $ t \in [0,T]$. There exists a unique positive solution $(\bar{\xi}_1,\bar{\xi}_2)$ such that $\bar{\xi}_2=h_1(\bar{\xi}_1)=h_2(\bar{\xi}_1)$.
\end{proposition}

By the previous analysis, for each $t \in [0,T]$, we can obtain a unique equilibrium solution of $(\bar{\xi}_1,\bar{\xi}_2)$ regardless of their constraints.
However, large values of $\bar{\xi}_1 $ and $ \bar{\xi}_2$ might lack practical meaning because they represent the amount of risk loading added into reinsurance premiums.
In our second attempt, we impose the constraints of $\xi_1(t)$ and $\xi_2(t)$ introduced in Definition \ref{admission} and then solve the equilibrium reinsurance premium strategy $(\xi_1^*(\cdot),\xi_2^*(\cdot))$. Specifically, we assume that $\xi_1(t) \in [\theta \beta,\eta \beta]$ and $\xi_2 \in [\theta,\eta]$.

\begin{proposition} \label{repro}
Assume that $Y_i \sim \exp(\beta) $, $i=1,2,...$, $\xi_1(t) \in [\theta \beta, \eta\beta]$, and $\xi_2(t) \in [\theta, \eta]$.
The equilibrium reinsurance premium strategies for two reinsurers are
\begin{align*}
         \xi_1^*(t) & =  \max \left \{ \theta \beta , \, \min \left \{ h_1^-(\xi_2^*(t)) ,\, \eta \beta\right \} \right \}, \\
         \xi_2^*(t) &= \begin{cases}   \max\left \{ \theta, \, \min \left \{ h_2(\theta \beta) , \, \eta   \right \} \right \}  , & \text{ if }   \bar{\xi}_1(t) <\theta \beta \\ \max\left \{ \theta, \, \min \left \{ \bar{\xi}_2 (t),\, \eta \right \} \right \}  , & \text{ if }  \theta \beta \leq \bar{\xi}_1(t) \leq \eta \beta \\ \max \left \{ \theta, \,  \min \{ h_2(\eta \beta), \eta \} \right \} , & \text{ if }   \eta \beta <\bar{\xi}_1(t)  .\end{cases}
\end{align*}
\end{proposition}

We compare the equilibrium solution without constraints with the boundary for the strategies. We can roughly understand the formula in the following way. When the equilibrium solution $\bar{\xi}_1(t)$ for reinsurer 1 exceeds $\eta \beta$, the maximal reinsurance premium that reinsurer 1 should charge is $\eta \beta$.
Reinsurer 2's corresponding strategy is $h_2(\eta \beta)$, which maximizes reinsurer 2's objective function given reinsurer 1's strategy $\eta \beta$. Considering the limitation of $\xi_2(t)$, reinsurer 2 can only set its relative security loading strategy between $\theta$ and $\eta$. To optimize its objective, the premium strategy for reinsurer 2 is $\max \left \{ \theta, \,  \min \{ h_2(\eta \beta), \eta \} \right \}$. 
Then, reinsurer 1 decides its strategy again, which is $\max \left \{ \theta \beta , \, \min \left \{ h_1^-(\max \left \{ \theta, \,  \min \{ h_2(\eta \beta), \eta \} \right \}) ,\, \eta \beta\right \} \right \}$. 
Now, $(\xi_1^*(t),\xi^*_2(t))$ can satisfy \eqref{equilibriumxi}, and they reach an equilibrium at time $t$.
When the equilibrium solution $\bar{\xi}_1(t)$ satisfies $\theta \beta \leq \bar{\xi}_1(t) \leq \eta \beta $, then we only need to consider the constraint for reinsurer 2's corresponding strategy. Therefore, reinsurer 2's premium strategy would be $\max\left \{ \theta, \, \min \left \{ \bar{\xi}_2 (t),\, \eta \right \} \right \} $. 
To reach equilibrium, reinsurer 1 will change its premium strategy into $\max \left \{ \theta \beta , \, \min \left \{ h_1^-(\max\left \{ \theta, \, \min \left \{ \bar{\xi}_2 (t),\, \eta \right \} \right \} ) ,\, \eta \beta\right \} \right \}$. 
In the last case, when $\bar{\xi}_1(t)<\theta \beta$, the best strategy that reinsurer 1 can choose is $\theta \beta$. Therefore, reinsurer 2's corresponding strategy is $ \max\left \{ \theta, \, \min \left \{ h_2(\theta \beta) , \, \eta   \right \} \right \} $. 
Then, reinsurer 1 decides her strategy again to reach equilibrium, which is $ \max \left \{ \theta \beta , \, \min \left \{ h_1^-( \max\left \{ \theta, \, \min \left \{ h_2(\theta \beta) , \, \eta   \right \} \right \}) ,\, \eta \beta\right \} \right \}$. 

\section{Numerical results and discussion}\label{numerical}
In this section, we provide a numerical example for the theoretical results we obtained in section \ref{example}. We choose the parameter values $t=0$, $T=8$, $\alpha=1$, $\sigma=1$, $\theta=0.1$, $\eta=0.9$, $\lambda=1$, $\beta=1$, $x_I=1$, $x_{R1}=10$, $x_{R2}=10$, $\rho_I(t)=0.1$, $\rho_{R1}(t)=0.1$, $\rho_{R2}(t)=0.1$, $\gamma_I=0.1$, $\gamma_{R1}=0.1$, and $\gamma_{R2}=0.1$. Therefore, we will provide the numerical results of equilibrium strategies for the three parties at time $0$.
It is worth noting that the initial asset of the reinsurer is set to be 10 times as much as that of the insurer to distinguish the reinsurers' and the insurer's solvency abilities in the market. We can also see that the safe loading for two reinsurers is limited in $[0.1,0.9]$.
Next, we will draw a phase diagram to illustrate the equilibrium strategies of three participants. Finally, we allow $\rho_I(t)$, $\rho_{R1}(t)$, $\rho_{R2}(t)$, $\gamma_I$, $\gamma_{R1}$, and $\gamma_{R2}$ to be free parameters and fix the rest to examine the sensitivity of the equilibrium strategies.

\subsection{The equilibrium strategies}

\begin{figure}[H]
\centering
\includegraphics[scale=0.5]{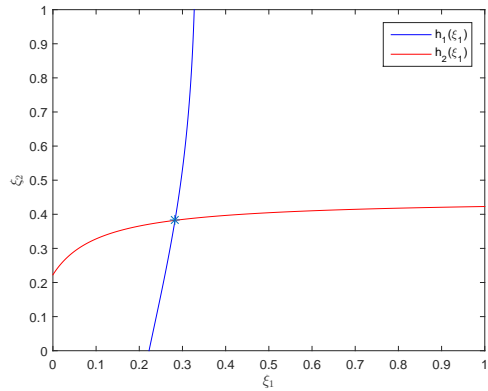}
\caption{The phase diagram of $\xi_1$ and $\xi_2$}
\label{fig:phase}
\end{figure}

\begin{figure}[H]
\centering
\includegraphics[scale=0.5]{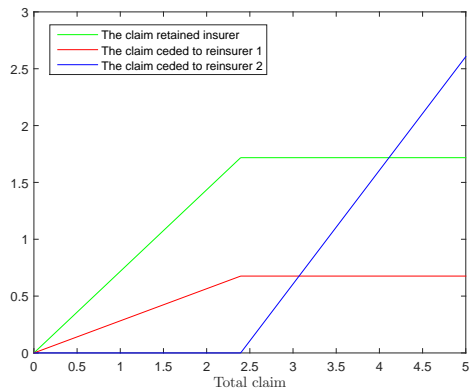}
\caption{The equilibrium reinsurance strategy}
\label{fig:reinsurance}
\end{figure}

We can obtain the phase diagram by plotting the functions $h_1(\xi_1)$ and $h_2(\xi_1)$ in Figure \ref{fig:phase}, where the intersection point of two curves is the equilibrium reinsurance premium strategy. We clearly find from the phase diagram that the equilibrium reinsurance premium strategies of reinsurer 1 and reinsurer 2 at time $0$ are $\xi_1^*(0)=0.28269$ and $\xi_2^*(0)=0.38225$, respectively. The reinsurance plan of the insurer is illustrated in Figure \ref{fig:reinsurance}. If the total claim is less than $d=2.3936$, the insurer retains 71.75\% of the total claim, and reinsurer 1 covers 28.25\%. If the total claim exceeds $d=2.3936$, the indemnity paid by insurer is at most 1.7175, and reinsurer 1 covers 0.6761 of the total claim. The excess of loss reinsurance signed with reinsurer 2 will pay all of the remaining indemnity.

\begin{figure}[htbp]
\centering
\begin{minipage}[t]{0.4\textwidth}
\centering
\includegraphics[width=7cm]{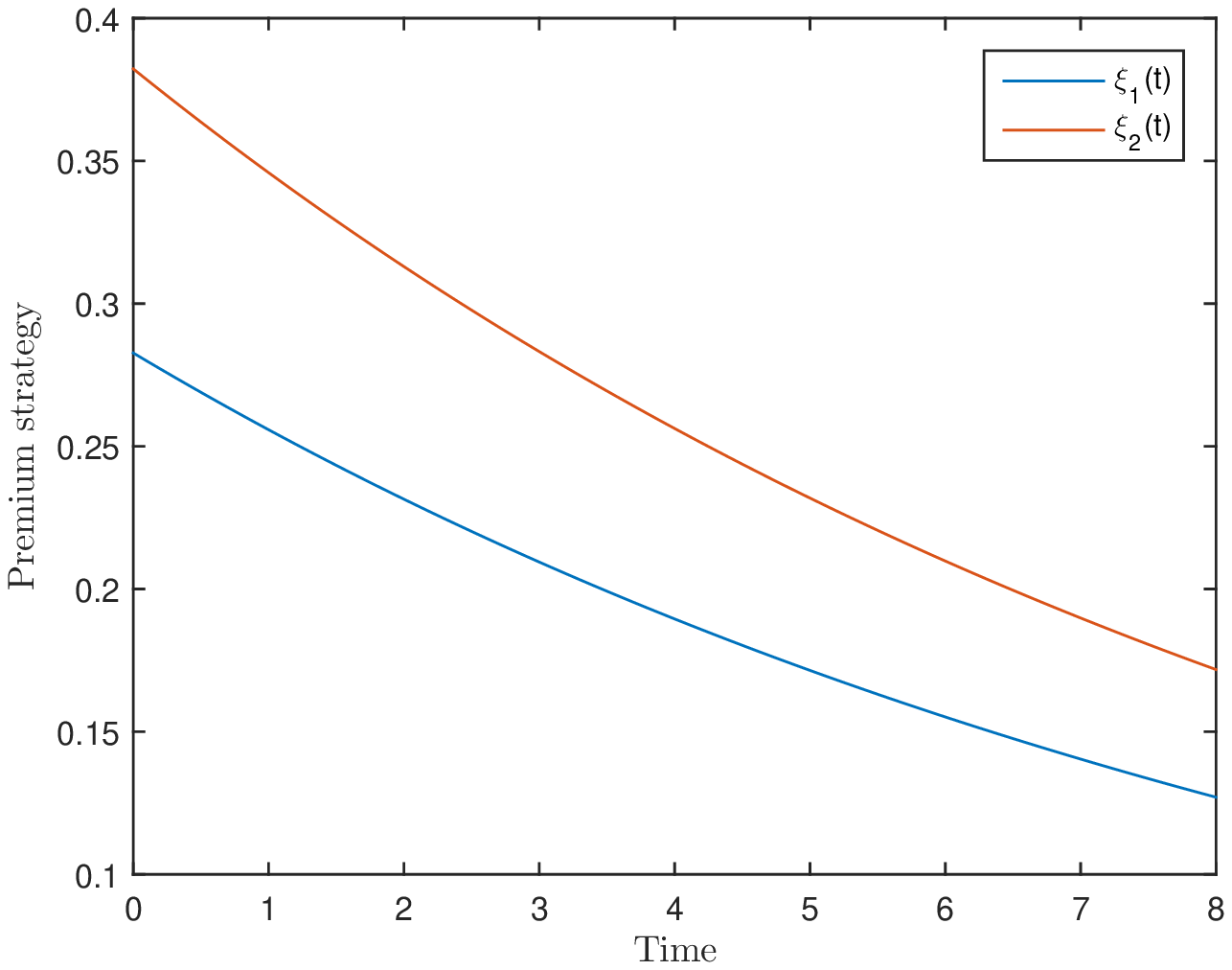}
\caption{Equilibrium reinsurance premium strategy}
\label{premium-time}
\end{minipage}
\begin{minipage}[t]{0.4\textwidth}
\centering
\includegraphics[width=7cm]{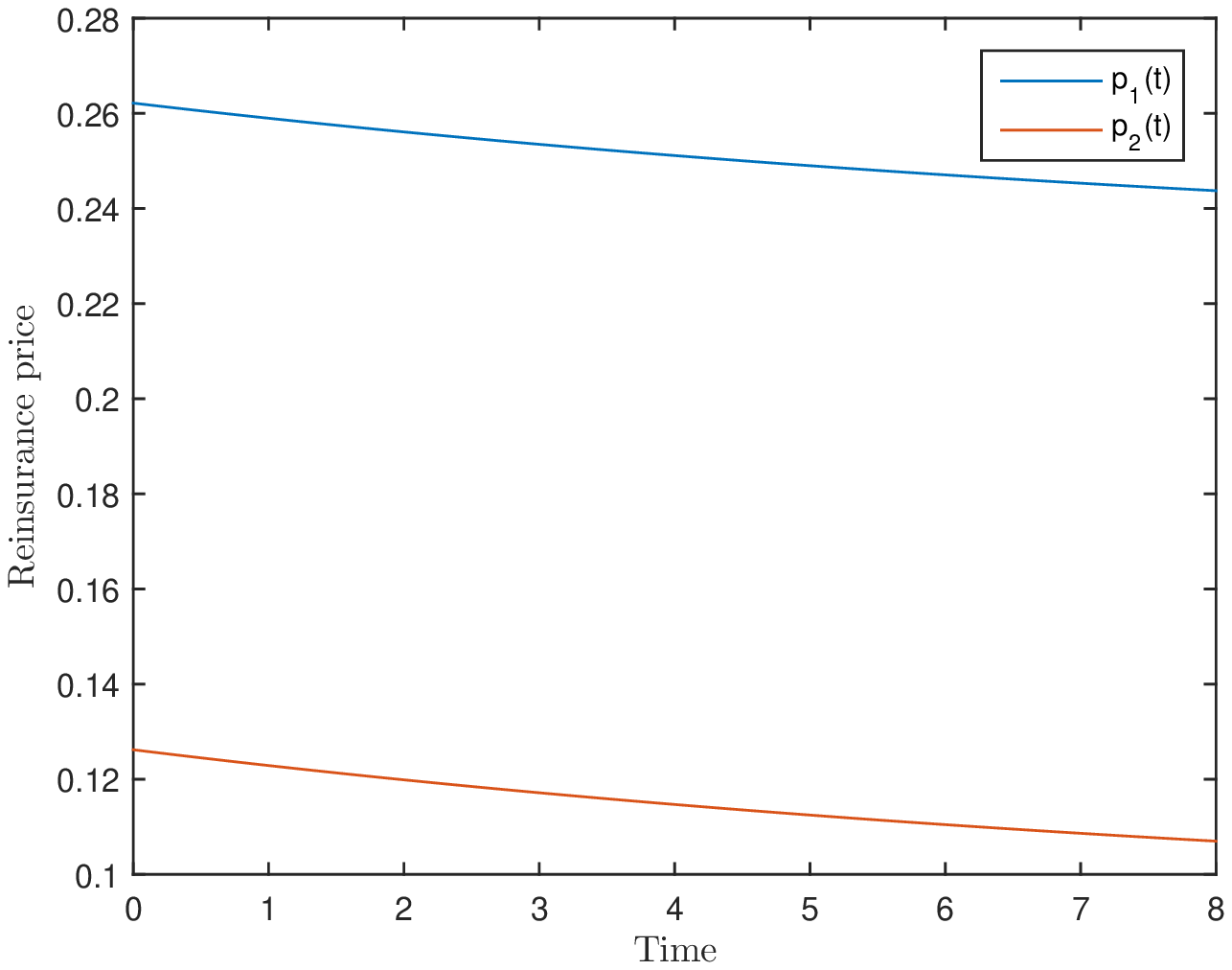}
\caption{Reinsurance price}
\label{price-time}
\end{minipage}
\end{figure}

In Figure \ref{premium-time}, we provide the trajectory of the equilibrium reinsurance premiums $\xi_1^*(t)$ and $\xi_2^*(t)$. As shown in the figure, the premium strategies for two reinsurers $\xi_1(t)$ and $\xi_2(t)$ both decrease over time. It is reasonable that the
reinsurers charge less to cover the risk as the uncertainty of the policy is reduced over time. Moreover, the prices for the two reinsurance policies do not change substantially over the time period according to Figure \ref{price-time}. For the policy with an 8-year lifetime and for one unit of claim $Y$, the price of the proportion reinsurance ranges between 0.2621 and 0.2437, and the price of the excess of loss reinsurance varies from 0.1262 to 0.1070.
The proportion ceded to reinsurer 1 is quite stable at approximately 0.2825, and the deductible is almost equal to 2.3936 from time $0$ to $8$. This means that the reinsurance policies signed with two reinsurers do not need to change considerably over time.

\subsection{Sensitivity analysis}

{
In this section, we discuss the effect of the expectation of claim, denoted by $\mu:=1/\beta$, risk attitude for each company ($\gamma_I$, $\gamma_{R1}$ and $\gamma_{R2}$) and credit interest rate ($\rho_I$, $\rho_{R1}$ and $\rho_{R2}$) on the equilibrium strategies. As the equilibrium strategies do not change substantially over time, we only consider the equilibrium strategies for two reinsurers at times $0$, $\xi_1^*(0)$ and $\xi_2^*(0)$. To simplify the expression, we use $\xi^*_1$ and $\xi^*_2$ to represent $\xi^*_1(0)$ and $\xi^*_2(0)$, respectively.

W first analyse the effect of $ \mu$ which is the expectation of a claim.
In this section, we denote by $ (\hat{\xi}_1, \hat{\xi}_2) $ the intersection point of functions $ h_1 $ and $ h_2 $.
Figure \ref{expectation-xi} shows the change of $\hat{\xi}_1 $ and $\hat{\xi}_2$ with respect to  $\mu$. Generally, $ \hat{\xi}_1  $ (resp.~$\hat{\xi}_2$) decreases (resp.~increases) with $\mu$ in a non-strictly sense.
Meanwhile, given a value of $ \mu $, the admissible range for $(\xi_1^* ,\xi_2^* )  $ is $ [0.1/\mu, 0.9 /\mu] \times [0,1,0.9]$.
 In what follows, we analyze the optimal solution according to the range of $ \mu $.
\begin{enumerate}
\item[1.] For  $0.3537\le \mu \le 2.3546$, i.e., $0.4247\le \beta \le 2.8269$, the safe loading constraint is not binding because $ (\hat{\xi}_1, \hat{\xi}_2) \in  [0.1/\mu, 0.9 /\mu] \times [0,1,0.9]$. Therefore, $ \xi_i^* = \hat{\xi}_i$, $i=1,2$, and  the equilibrium solution lies in the admissible region as show in Figure \ref{fig:phase}.
By substituting \eqref{xi1-2-2} into \eqref{xi1-2-1}, we can see that $\xi_1^*$ does not depend on $\beta$.  It implies $\xi_1^*=0.28269$.  In addition, $\xi_2^*$ decreases with $\beta$ according to \eqref{xi1-2-2}, which lead to $\xi_2^*$ increases with $\mu$.
\item[2.] If $\mu\ge 3.1837$, i.e., $\beta \le 0.3141$, as shown in Figure \ref{beta02}, the intersection point $ (\hat{\xi}_1, \hat{\xi}_2)  $ is outside and upper-right to the admissible region. Consequently, the equilibrium solution is the upper-right corner of admissible region, that is  $  (\xi_1^* ,\xi_2^* ) = (\beta\eta,\eta) = (0.9/\mu , 0.9)$.
\item[3.] For  $2.3546 \le \mu \le 3.1837$, i.e., $0.3141 \le \beta \le 0.4247$,  the intersection point $ (\hat{\xi}_1, \hat{\xi}_2)  $ is outside the admissible region.
Precisely, $ \hat{ \xi}_1 \in [0.1/\mu, 0.9/\mu] $ satisfies the safe loading condition, while $ \hat{ \xi}_2 > 0.9 $ does not. Therefore, we have to limit $ \xi_2^* = 0.9 $ in the equilibrium solution. It follows that  the equilibrium solution is on the boundary of the admissible region, as shown in Figure \ref{beta04}.
\item [4.] If  $0.2525 \le \mu \le 0.3537$, i.e., $ 2.8269\le \beta \le 3.9610$,   the intersection point $ (\hat{\xi}_1, \hat{\xi}_2)  $ is outside the admissible region.
Precisely $ 0.1 \leq \hat{ \xi}_2 \leq 0.9 $ satisfies the condition, while $ \hat{\xi}_1  < 0.1/\mu$ does not.
Thus, the equilibrium point $(\xi_1^*, \xi_2^*)$ is given by $ \xi_1^* =\beta\theta = 0.1 / \mu $ and $ \xi_2^* = h_2(\xi_1^*)$
 as shown in Figure \ref{beta3}.
 \item[5.] In the last case where $\mu \le 0.2525$, i.e., $\beta \ge 3.9610$,   the intersection point $ (\hat{\xi}_1, \hat{\xi}_2)$ lies in the lower left of the admissible region. Therefore,  the equilibrium solution is $(\xi_1^*, \xi_2^*) = (\beta\theta, \theta) = (0.1/\mu, 0.1)$, as shown in Figure \ref{beta45}.
\end{enumerate}

Figure $\ref{expectation-qd}$ shows the effect of $\mu$ on the equilibrium reinsurance strategy, which is the proportion $q$ (see left $y$ axis) ceded to reinsurers 1 and deductible $d$ (see right $y$ axis) for reinsurer 2.
The proportion ceded to reinsurers is decreasing with $\xi_1$; thus, $q$ is increasing with $\mu$. The proposition rises sharply from $1$ to $28.25\%$ before $\mu=0.3537$ and then goes to a platform. Finally, it rises slowly to 1 as $\mu \to +\infty$. Because $d$ is increasing with $\xi_2$ and decreasing with $\xi_1$, the deduction is also increasing as the claim risk increases. As $\mu \to \infty$, $d \to +\infty$.

\begin{figure}[htbp]
\centering
\begin{minipage}[t]{0.48\textwidth}
\centering
\includegraphics[width=6cm]{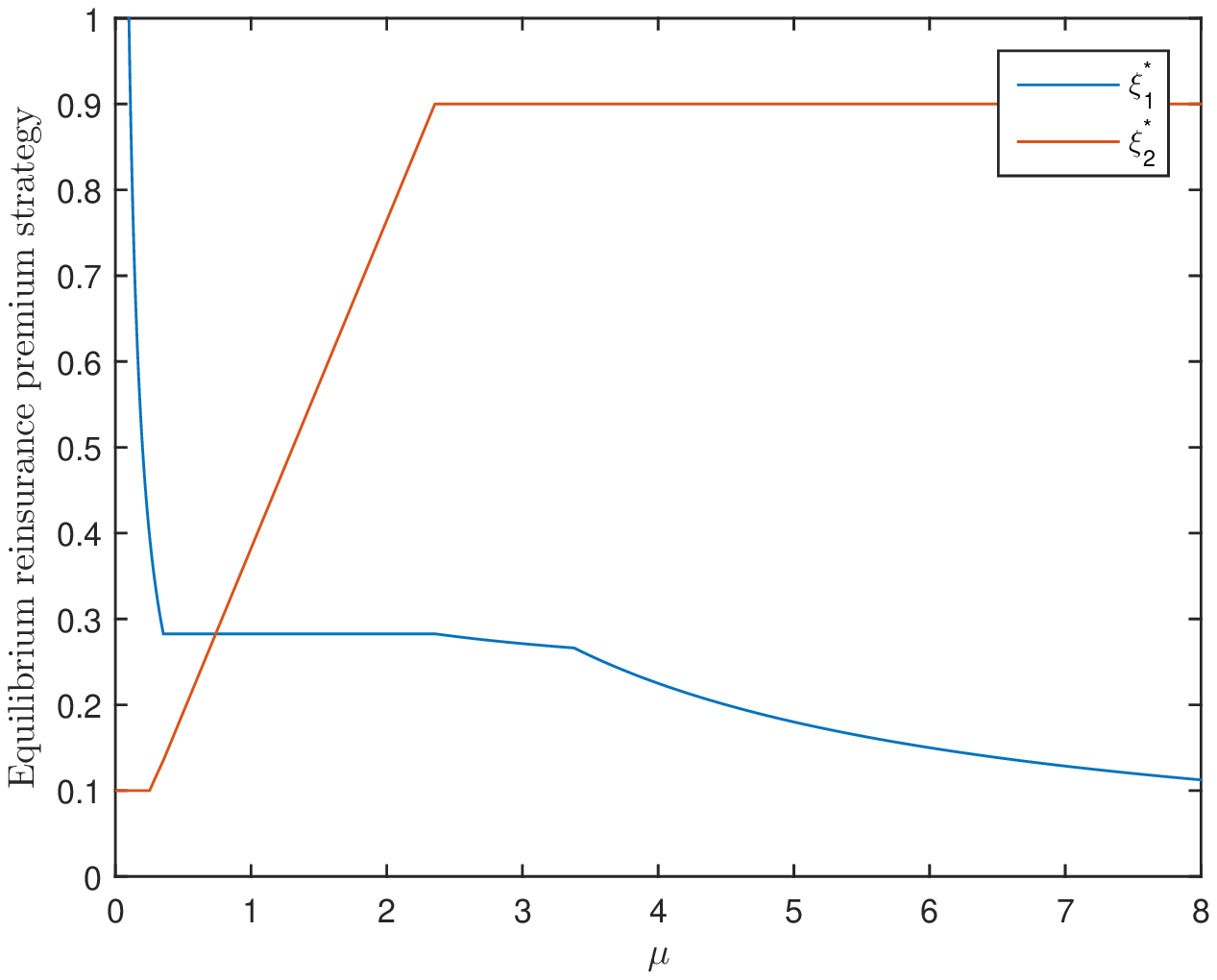}
\caption{Effect of $\mu$ on $\xi_1^*$ and $\xi_2^*$}
\label{expectation-xi}
\end{minipage}
\begin{minipage}[t]{0.48\textwidth}
\centering
\includegraphics[width=6cm]{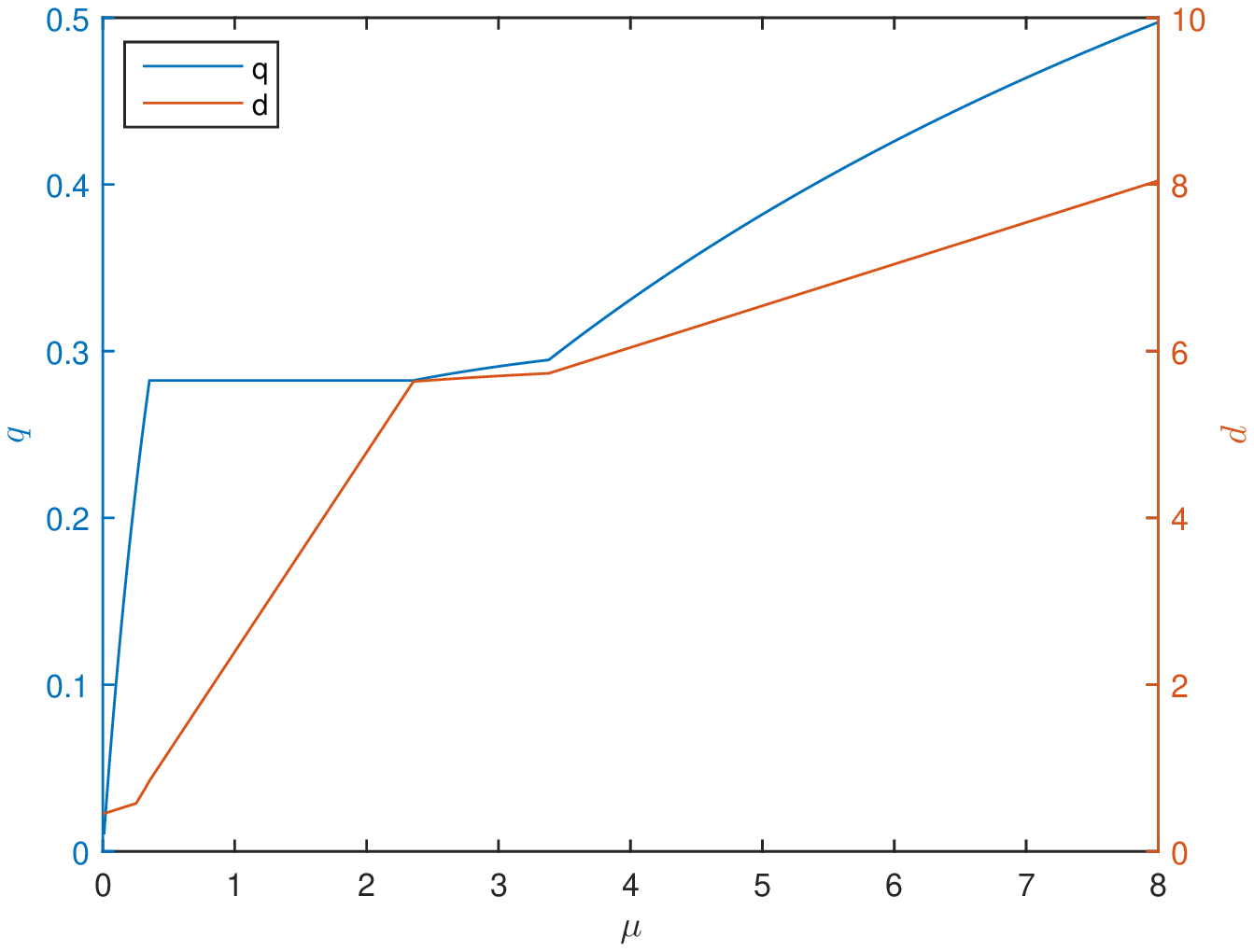}
\caption{Effect of $\mu$ on $q$ and $d$}
\label{expectation-qd}
\end{minipage}
\end{figure}

\begin{figure}[htbp]
\centering
\begin{minipage}[t]{0.48\textwidth}
\centering
\includegraphics[width=6cm]{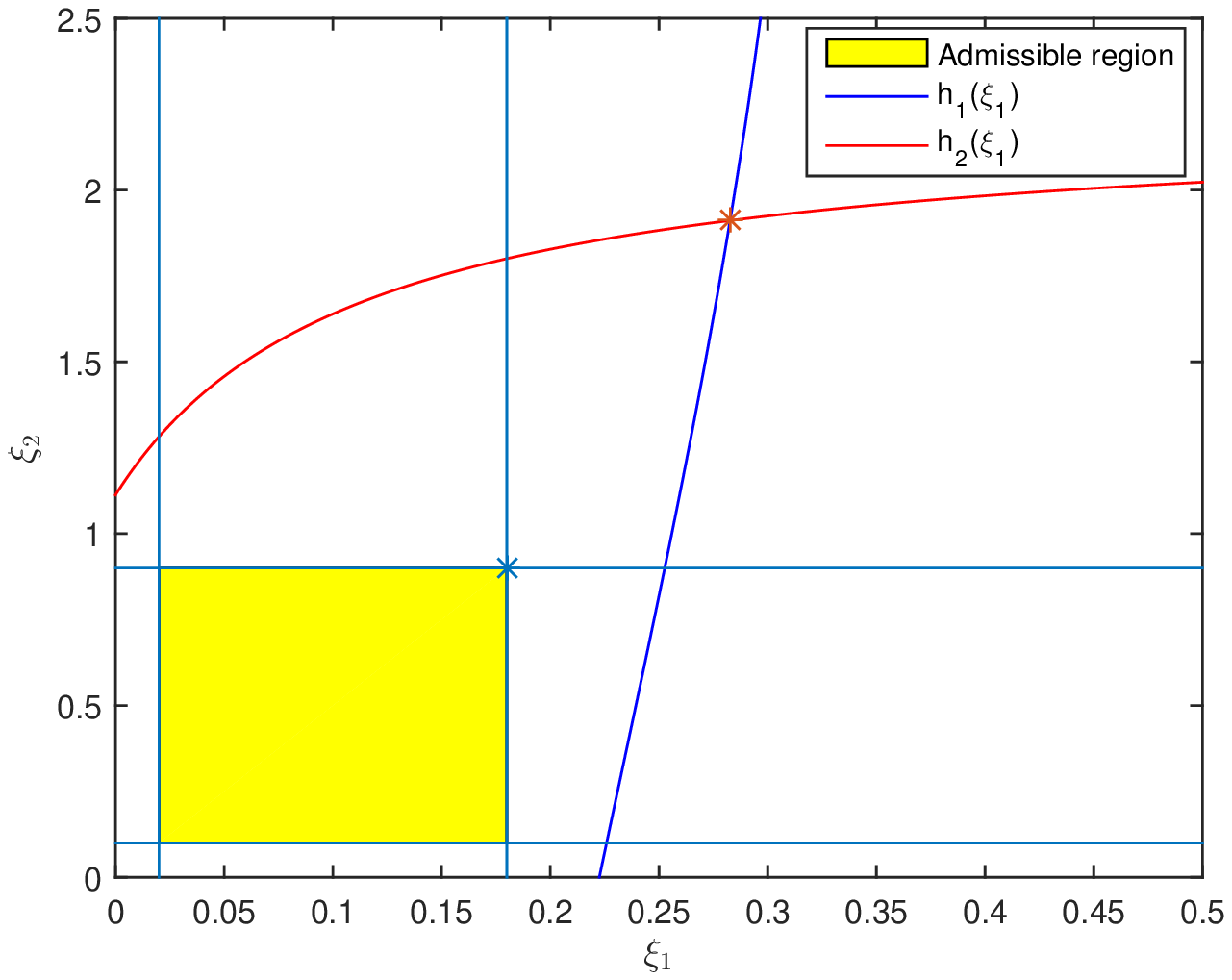}
\caption{Phase diagram of $\beta=0.2$}
\label{beta02}
\end{minipage}
\begin{minipage}[t]{0.48\textwidth}
\centering
\includegraphics[width=6cm]{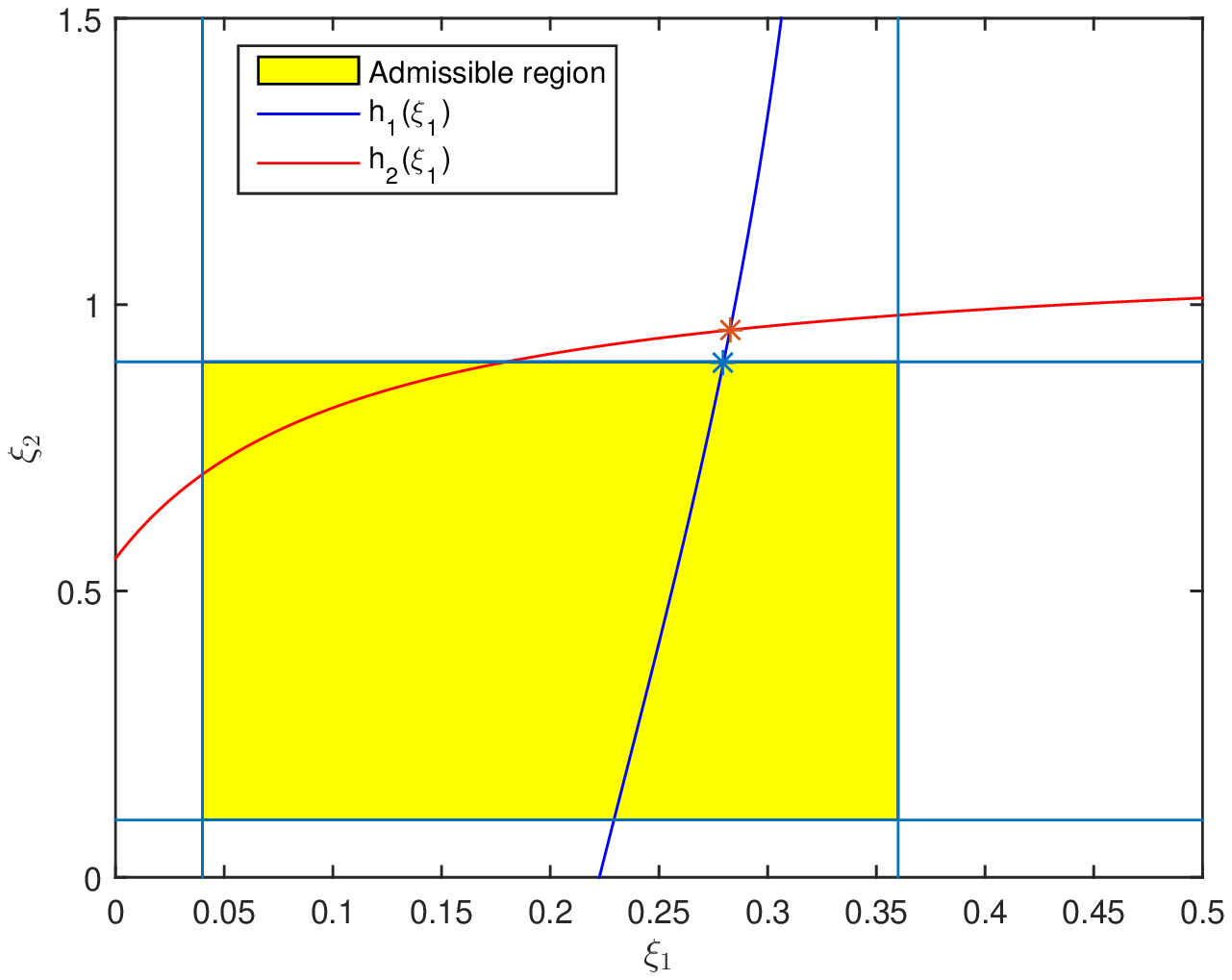}
\caption{Phase diagram of $\beta=0.4$}
\label{beta04}
\end{minipage}
\centering
\begin{minipage}[t]{0.48\textwidth}
\centering
\includegraphics[width=6cm]{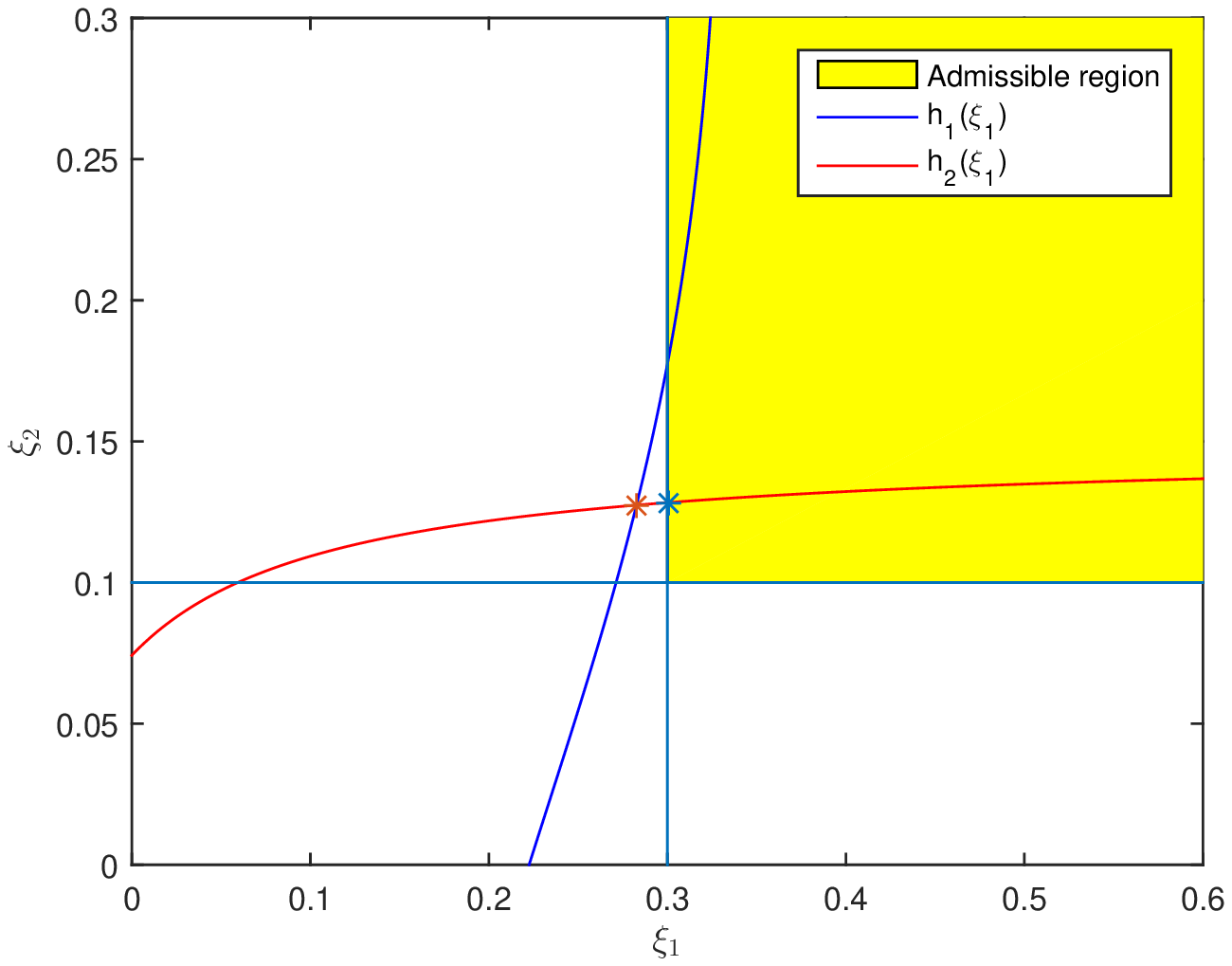}
\caption{Phase diagram of $\beta=3$}
\label{beta3}
\end{minipage}
\begin{minipage}[t]{0.48\textwidth}
\centering
\includegraphics[width=6cm]{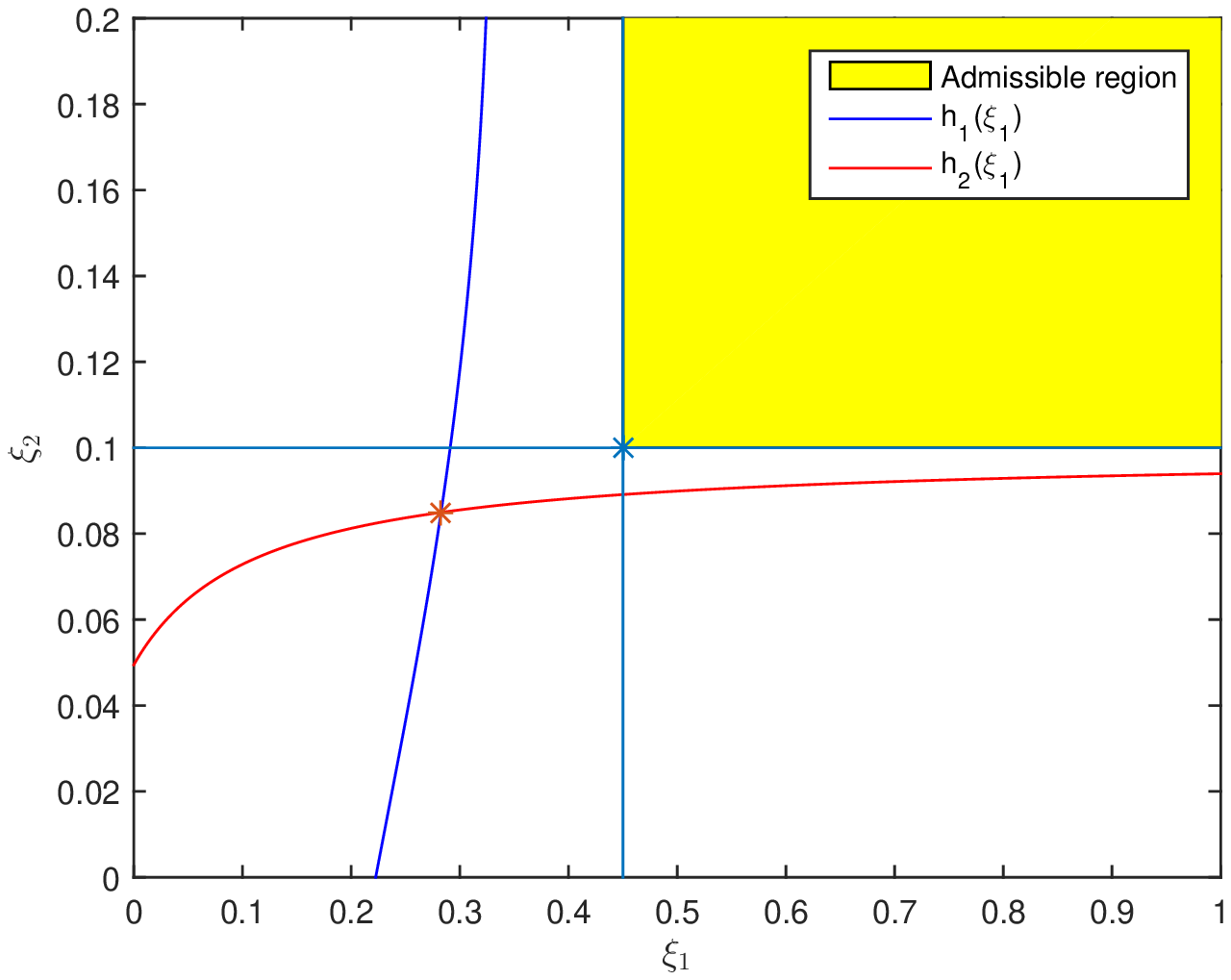}
\caption{phase diagram of $\beta=4.5$}
\label{beta45}
\end{minipage}
\end{figure}

\begin{figure}[htbp]
\centering
\begin{minipage}[t]{0.48\textwidth}
\centering
\includegraphics[width=6cm]{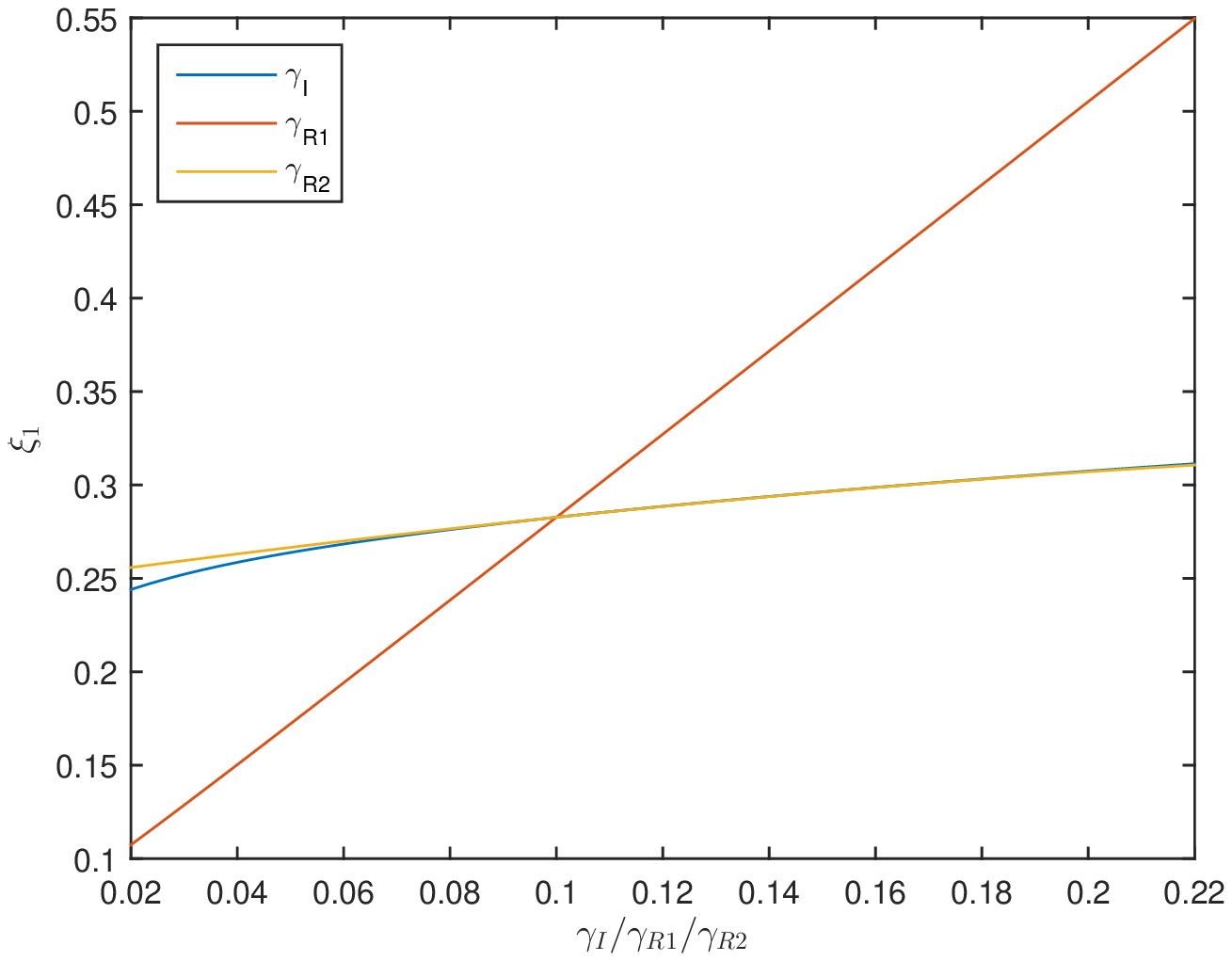}
\caption{Effect of risk attitudes on $\xi_1^*$}
\label{gamma-xi1}
\end{minipage}
\centering
\begin{minipage}[t]{0.48\textwidth}
\centering
\includegraphics[width=6cm]{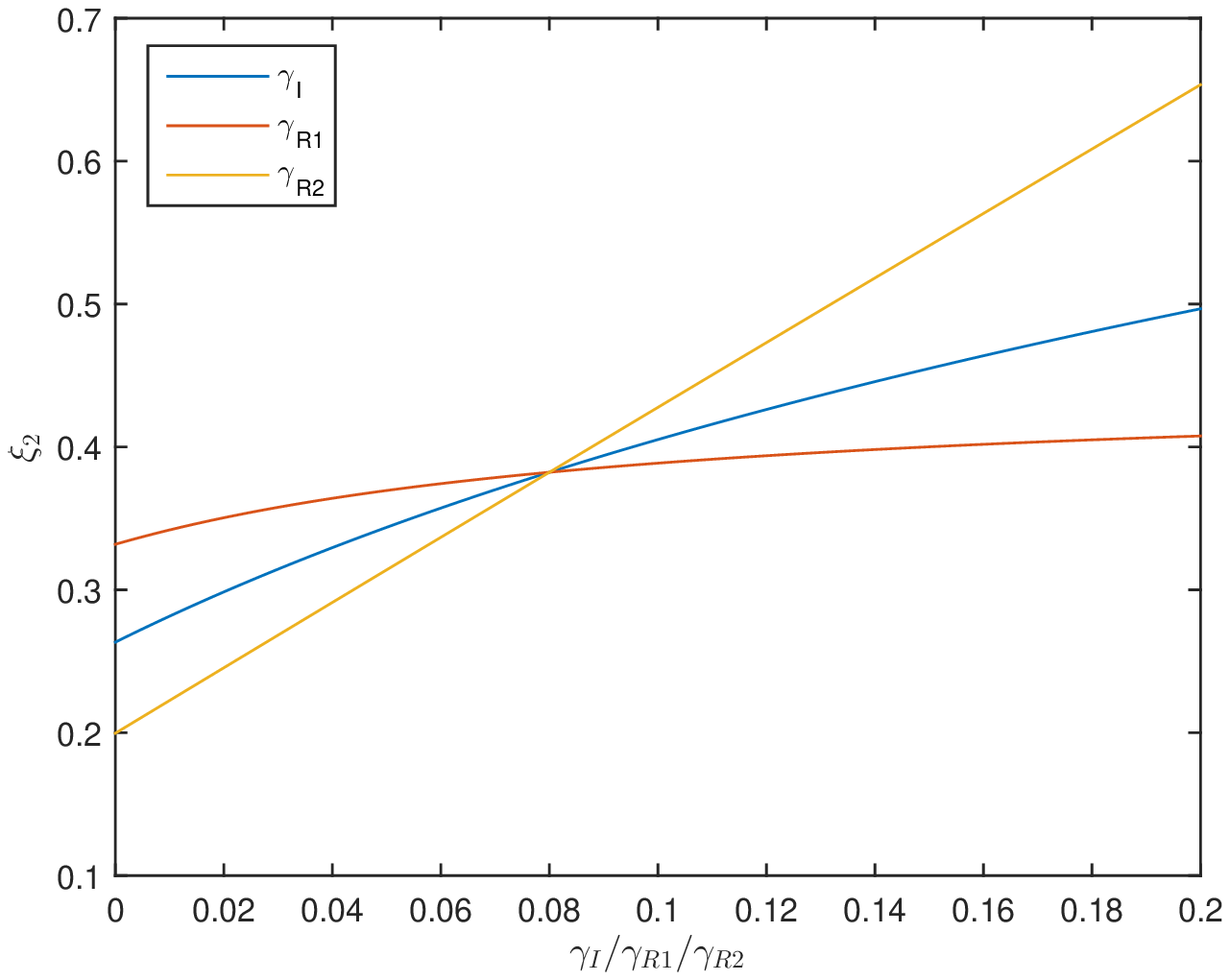}
\caption{Effect of risk attitudes on $\xi_2^*$}
\label{gamma-xi2}
\end{minipage}
\centering
\begin{minipage}[t]{0.48\textwidth}
\centering
\includegraphics[width=6cm]{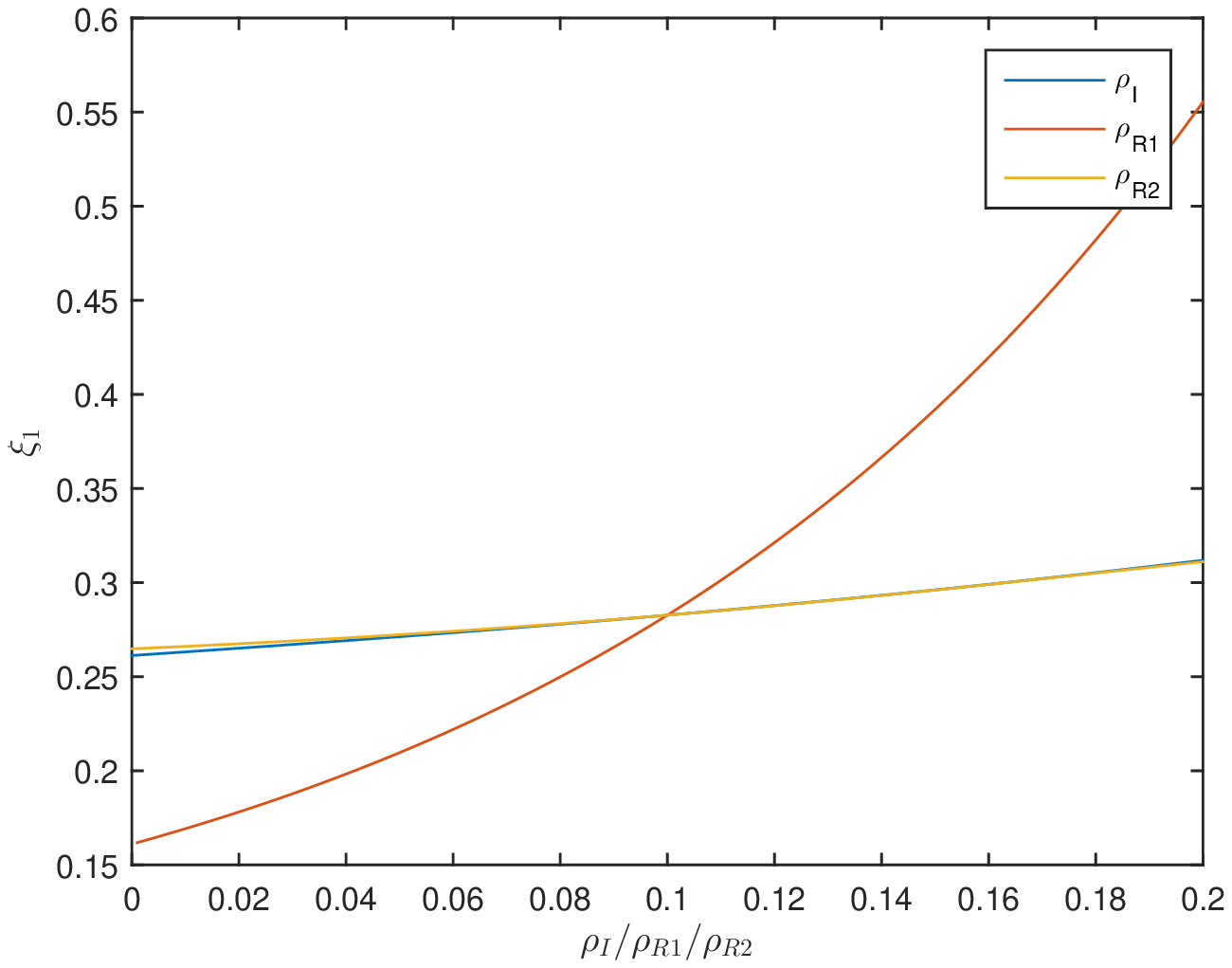}
\caption{Effect of interest rate on $\xi^*_1$}
\label{rho-xi1}
\end{minipage}
\centering
\begin{minipage}[t]{0.48\textwidth}
\centering
\includegraphics[width=6cm]{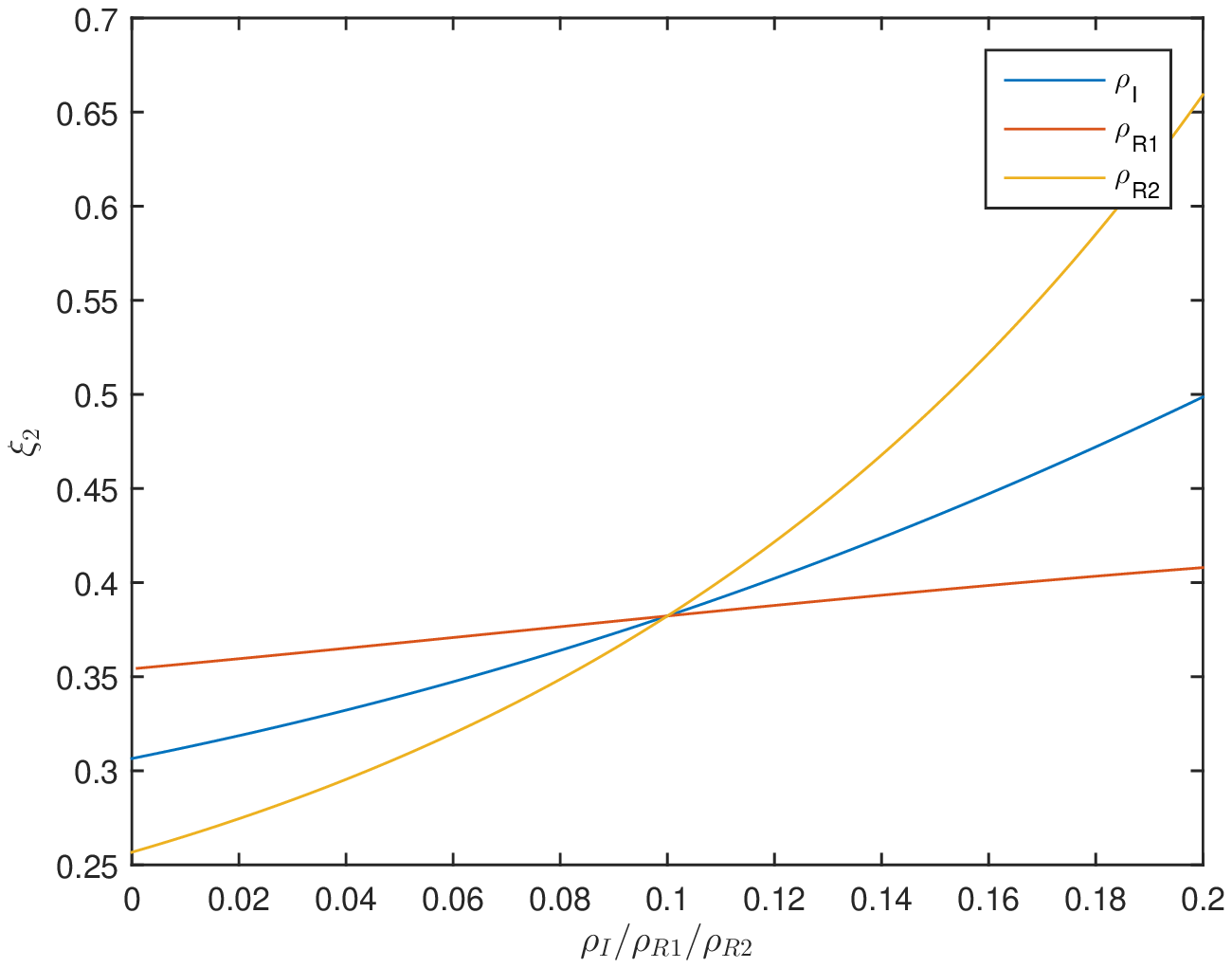}
\caption{Effect of risk attitudes on $\xi_2^*$}
\label{rho-xi2}
\end{minipage}
\end{figure}

\begin{figure}[htbp]
\centering
\begin{minipage}[t]{0.48\textwidth}
\centering
\includegraphics[width=6cm]{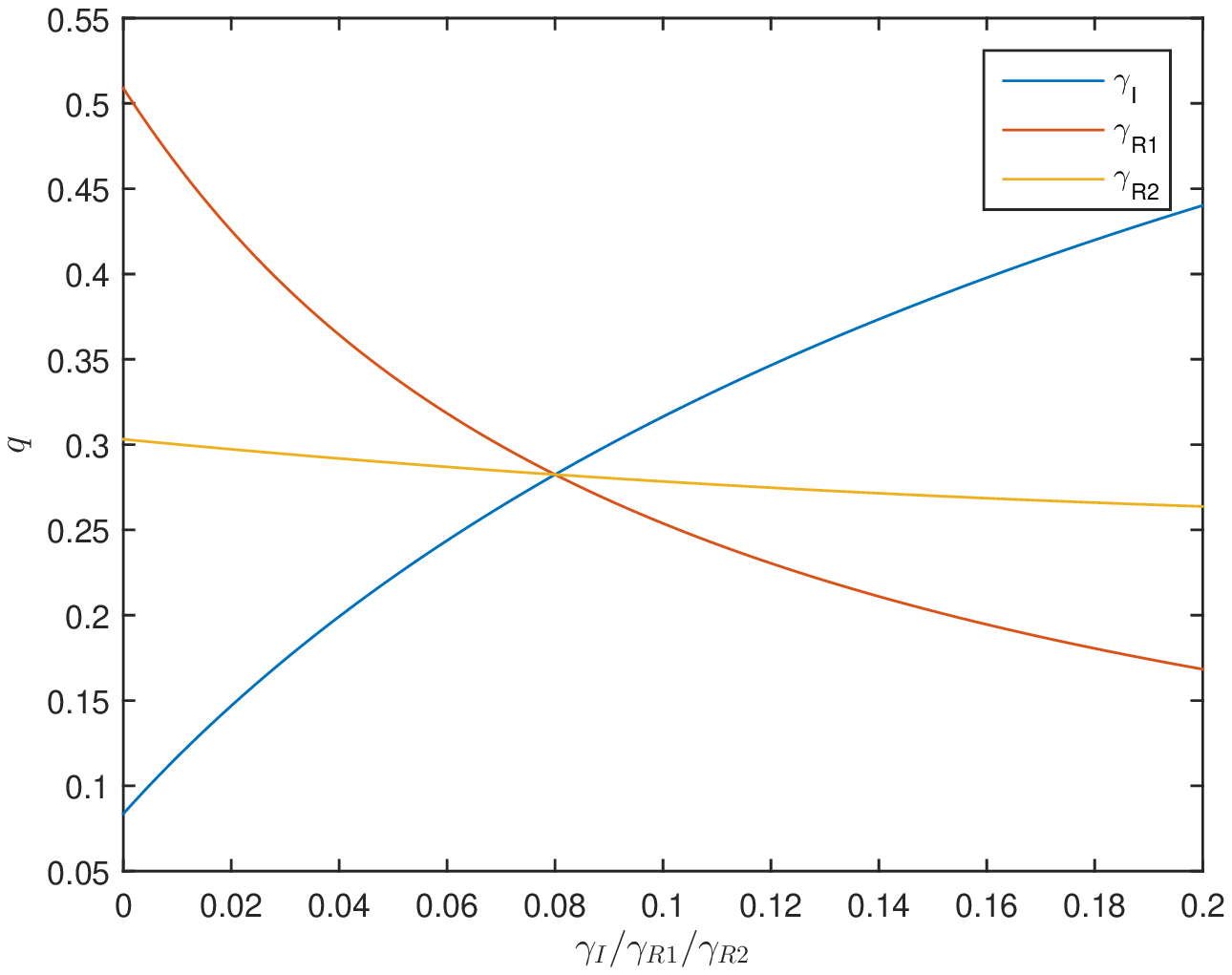}
\caption{Effect of risk attitude on $q$}
\label{gamma-q}
\end{minipage}
\centering
\begin{minipage}[t]{0.48\textwidth}
\centering
\includegraphics[width=6cm]{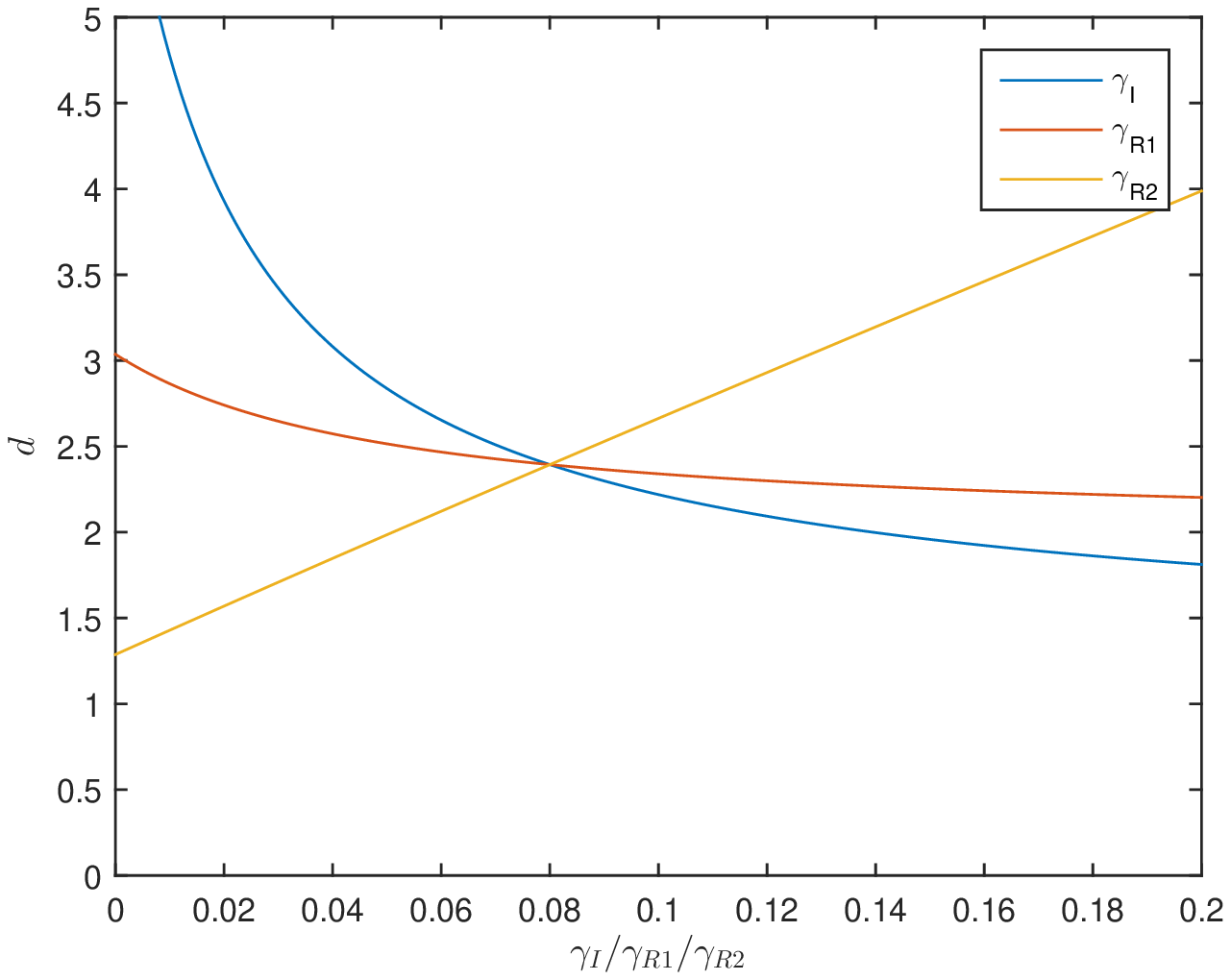}
\caption{Effect of risk attitude on $d$}
\label{gamma-d}
\end{minipage}
\centering
\begin{minipage}[t]{0.48\textwidth}
\centering
\includegraphics[width=6cm]{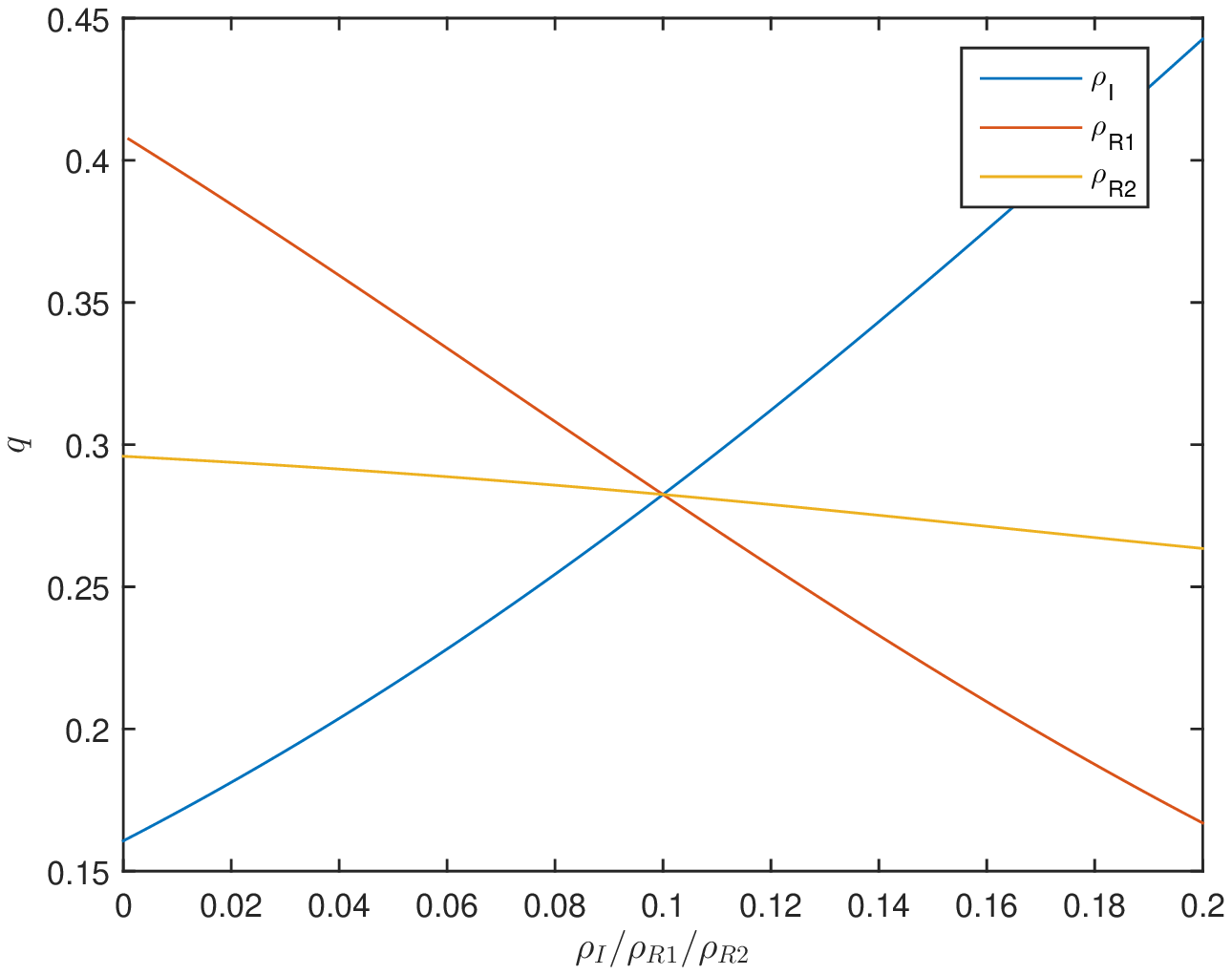}
\caption{Effect of interest rate on $q$}
\label{rho-q}
\end{minipage}
\centering
\begin{minipage}[t]{0.48\textwidth}
\centering
\includegraphics[width=6cm]{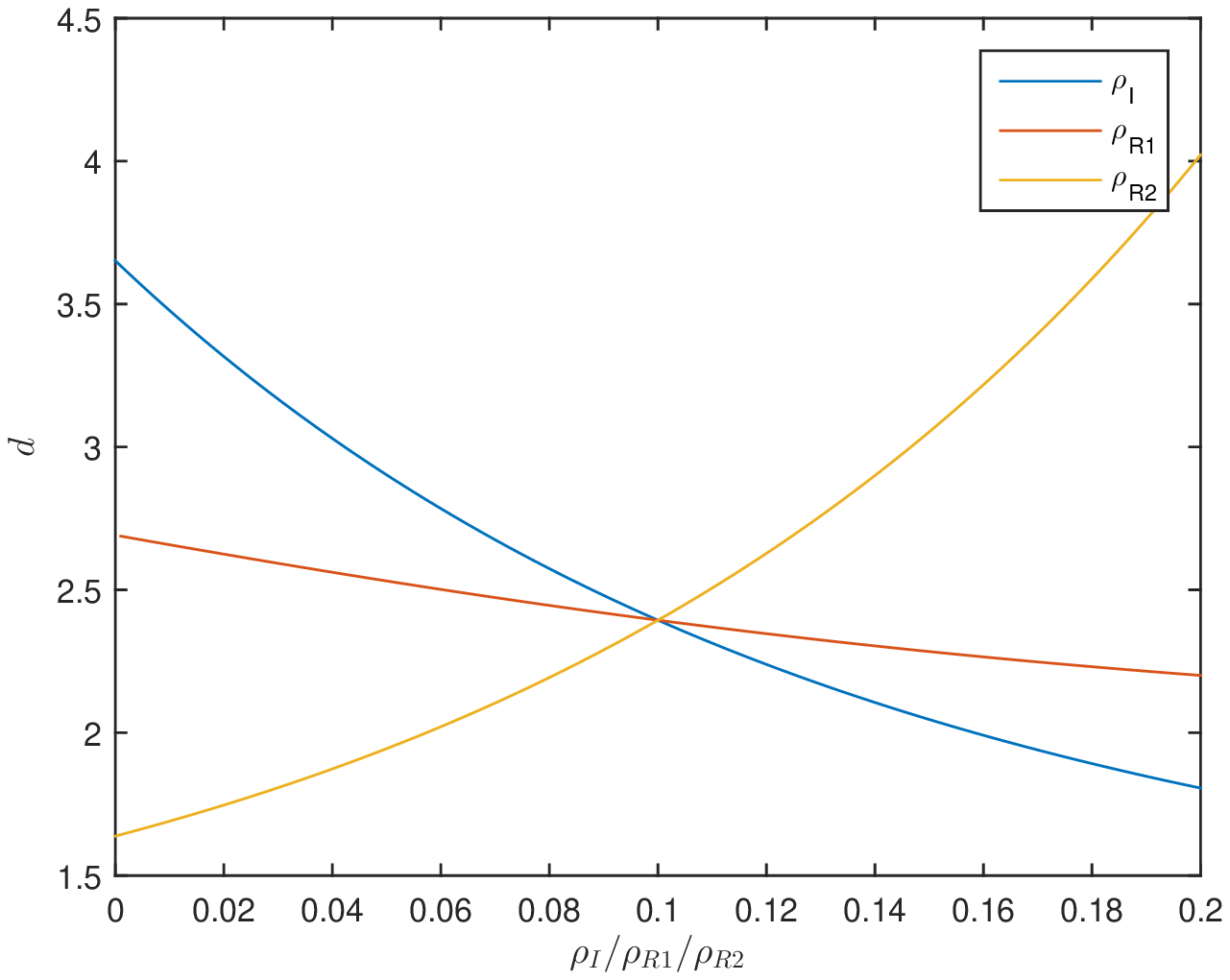}
\caption{Effect of interest rate on $d$}
\label{rho-d}
\end{minipage}
\end{figure}

Next,  we show the how the risk attitudes ($\gamma_I$, $\gamma_{R1}$ and $\gamma_{R2}$) and interest rates ($\rho_I$, $\rho_{R1}$ and $\rho_{R2}$) affect the equilibrium reinsurance ($q$ and $d$) and premium ($\xi_1^*$ and $\xi_2^*$) in Figure \ref{gamma-xi1} to \ref{rho-d}, in which the $x$ axis is used for the value of risk aversion degree or the value of interest for different companies. We state that when we analyse the effect of one parameter, we will keep the other parameters as the value we set in the basic case.
As the risk aversion and interest rate for each party always appear in the same place when we compute the equilibrium strategy, they will have the same effect on the equilibrium strategies.

Figures \ref{gamma-xi1} and \ref{gamma-xi2} illustrate that both $\xi_1^*$ and $\xi_2^*$ will increase no matter which company become more risk aversion. Furthermore, we can also see that the change of reinsurer 1's risk attitude has the greatest impact on $\xi_1^*$. And $\gamma_I$ and $\gamma_{R2}$ do not affect $\xi_1^*$ too much. Risk attitude of reinsurer 2 has greater impact on $\xi_2^*$ than insurance company and reinsurer 1. But $\gamma_I$ changes $\xi_2^*$ more than $\gamma_{R1}$. As the interest rates always appear together with corresponding risk attitude, the interest rates have the same effect as risk attitudes, which is shown in \ref{rho-xi1} to \ref{rho-xi2}. The increasing of risk attitude or interest rate will increase the cost of the insurance product or reinsurance product for companies. Therefore, for the insurance company, it has higher demand for reinsurance which leads to an increase in the reinsurance premium. For the reinsurance companies, they will reduce the supply for reinsurance which also lead to higher price for reinsurance in the market.

Figures \ref{gamma-q} to \ref{rho-d} show that the ceded proportion $q$ increases with $\gamma_I$ and $\rho_I$, decreasing with $\gamma_{R1}$, $\rho_{R1}$ and $\gamma_{R2}$, $\rho_{R2}$. And the deductible $d$ increases with $\gamma_{R2}$ and $\rho_{R2}$, decreasing with $\gamma_{R1}$, $\rho_{R1}$ and $\gamma_{I}$, $\rho_{I}$.
Furthermore, we can see that the risk attitude and interest rate affect both $q$ and $d$ strongly. On the other hand, the risk attitude and interest rate for reinsurance companies will affect their own product more. We can see that $\gamma_{R1}$ and $\rho_{R1}$ influence $q$ more, while $\gamma_{R2}$ and $\rho_{R2}$ influence $d$ more. For the same reason we mentioned before, the higher risk aversion and interest rate of insurance company increase the demand for reinsurance. Thus, $q$ increases and $d$ decreases. While the higher $\gamma_{R1}$ and $\rho_{R1}$ reduce the supply of proposition reinsurance, the demand for stop-loss reinsurance increases; thus, both $q$ and $d$ decreases. However, the higer $\gamma_{R2}$ and $\rho_{R2}$ will lead to greater $d$ and smaller $q$. It means that the reinsurance market will provide less reinsurance to insurance company in this case. Thus, the insurer needs to retain more risk itself.}

\section{Conclusion} \label{conclusion}

In this paper, we studied the stochastic differential reinsurance game between one insurer and two reinsurers under the Stackelberg game and Nash game. The insurer accepts risk from the insured with the expected value premium and cedes part of risk to two reinsurers who apply different premium principles. We study the reinsurance strategies of the insurer and reinsurance premium strategies of reinsurers under the mean-variance framework. The insurer will sign proposition reinsurance with reinsurer 1 who uses the variance premium principle and exceeds the loss reinsurance contract with reinsurer 2 who applies the expected value premium. The ceded proportion and deduction depend on the equilibrium reinsurance premium from the price competition between the two reinsurers. We also provided the solution of the equilibrium premium strategies of price competition when the claim size follows an exponential distribution through a phase diagram. The numerical analysis of this example shows how the equilibrium premium strategy changes with time, the claim size distribution, risk aversion and the interest rate, which can help us to understand the change in reinsurance price in the market. The model can be further extended to general cases with more insurers and more reinsurers.

\begin{appendices}
\section{Appendix} \label{appendix}
\subsection{Proof of Proposition \ref{pro: independent}}\label{proof-independ}
\begin{proof}
The proof follows from \cite{Basak2009Dynamic}. From the surplus process \eqref{wealth1}-\eqref{wealth3}, we calculate the objective function for the insurer and reinsurers as follows:
\begin{align}\label{proinde1}
& \, J_I( t, x_I ; l_1 , l_2 , \xi_1 , \xi_2 ) =  \E_{t,x_I} \left [X_I(T) \right ] -\frac{\gamma_I}{2} \Var_{t,x_I} \left ( X_I(T) \right ) \\
=& \, e^{\int_t^T \rho_I(u) du}x_I
 +\E_{t, x_I} \Bigl[ \int_t^T \int_{\R^+} e^{\int_s^T \rho_I(u) du} \Bigl[ y-l_1(s, y)-l_2(s,y) + \theta y -\xi_1(s)l^2_1(s,y) \notag \\
& \, - \xi_2 (s) l_2(s, y) -\frac{\gamma_I}{2} e^{\int_s^T \rho(u) du} (y-l_1(s,y)-l_2(s,y))^2 \Bigr] v(\sd y) \d s\Bigr ] \notag \\
& \,  - \frac{\gamma_I}{2} \Var_{t,x} \Bigl (  \int_t^T \int_{\R^+} e^{\int_s^T \rho_I(u) du} \Bigl[ (1+\theta) y-l_1(s, y)  - (1+\xi_2(s)) l_2(s, y) -\xi_1(s)l^2_1(s, y) \Bigr] v(\sd y) \d s \Bigr ) , \notag
\end{align}
\begin{align}\label{proinde2}
& \, J_{ R1}( t, x_{R1} ; l_1 ,\xi_1  ) = \E_{t, x_{R1}} \left [X_{R1}(T)\right] -\frac{\gamma_{R1}}{2} \Var_{t, x_{R1}} \left ( X_{R1}(T)\right )  \\
= & \, e^{\int_t^T \rho_{R1}(u) du}x_{R1}
 +\E_{t, x_{R1}} \left [ \int_t^T \int_{\R^+} e^{\int_s^T \rho_{R1}(u) du} \Bigl[l_1(s,y)+ \xi_1(s)l_1^2(s, y) - \frac{\gamma_{R1}}{2} e^{\int_s^T \rho_{R1}(u) du} l_1^2(s,y) \Bigr] v(\sd y)\d s \right ]  \notag \\
& \,  - \frac{\gamma_{R1}}{2} \Var_{t,x}  \left ( \int_t^T \int_{\R^+} e^{\int_s^T \rho_{R1}(u) du}[ l_1(s,y)+ \xi_1(s)l_1^2(s, y)] v(\sd y) \d s \right ), \notag
\end{align}
and
\begin{align}\label{proinde3}
& \, J_{ R2}( t, x_{R2} ;  l_2 , \xi_2) = \E_{t, x_{R2}} \left [X_{R2}(T) \right ] -\frac{\gamma_{R2}}{2} \Var_{t, x_{R2}} \left ( X_{R2}(T) \right )  \\
=&\, e^{\int_t^T \rho_{R2}(u) du}x_{R2}
  +\E_{t, x_{R2}} \left [ \int_t^T \int_{\R^+} e^{\int_s^T \rho_{R2}(u) du} \Bigl[(1+ \xi_2(s))l_2(s, y) - \frac{\gamma_{R2}}{2} e^{\int_s^T \rho_{R2}(u) du} l_2^2(s,y) \Bigr] v( \sd y)\d s \right  ] \notag \\
& \, - \frac{\gamma_{R2}}{2} \Var_{t,x} \left (  \int_t^T \int_{\R^+} e^{\int_s^T \rho_{R2}(u) du}(1+\xi_2(s))  l_2(s, y) v(\sd y )\d s \right ). \notag
\end{align}
As we can see, $\{l_1(\cdot),l_2(\cdot),\xi_1(\cdot), \xi_2(\cdot)\}$ only appears in the expectation and variance terms of the objection function, and those terms are independent of $x_I$, $x_{R1}$ and $x_{R2}$. Thus, their optimal trajectories must be state independent. Therefore, each player's objection function only depends on its own surplus and is independent of others'. The same for the value function.
\end{proof}

\subsection{Proof of Proposition \ref{inpro} }\label{proof-inpro}
\begin{proof}
For notational simplicity, denote $V^*_I(t,x_I)=V_I(t,x_I;\xi_1,\xi_2)$ and $g^*_I(t,x_I)=g(t,x_I;\xi_1,\xi_2)$. Combining this with \eqref{gen1}, the extended HJB equation for $V_I(t,x_I;\xi_1,\xi_2)$ in \eqref{sup1} is
\begin{align*}
& \mathrm{HJB}_I (t, x_I ;l_1, l_2 , \xi_1, \xi_2  ) \\
=&  \sup\limits_{\{l_1,l_2\} \in \mathscr{A}_I } \left \{  \mathscr{L}_I^{l_1,l_2,\xi_1,\xi_2}[V_I^* (t,x_I)]- \frac{\gamma_I}{2} \mathscr{L}_I^{l_1,l_2,\xi_1,\xi_2}[ (g_I^*)^2(t, x_I)]+\gamma_I g_I^* (t,x_I ) \mathscr{L}_I^{l_1,l_2,\xi_1,\xi_2}[g_I^* (t,x_I )] \right \}\\
=&\sup_{\{l_1,l_2\}\in \mathscr{A}_I}\bigg\{\mathscr{L}_I^{l_1,l_2,\xi_1,\xi_2}[V_I^*(t,x_I)]-\frac{\gamma_I}{2}\left [  2 g_I^*(t,x_I)\frac{\partial g_I^*(t,x_I)}{\partial t}+2g_I^*(t,x_I)\frac{\partial g_I^*(t,x_I)}{\partial x_I}\left ( c-p_1(t)-p_2(t)+\rho_I(t)x_I \right ) \right ]\\
&~~~~~~~~~~~~~~+ \frac{\gamma_I}{2} \int_{\R^+} (g_I^*)^2 ( t,x_I-y+l_1+l_2)- (g_I^*)^2(t,x_I)v(\sd v) \\
&~~~~~~~~~~~~~~ + \gamma_I g_I^*(t,x_I)\left [\frac{\partial g_I^*(t,x_I)}{\partial t}+\frac{\partial g_I^*(t,x_I)}{\partial x_I}\left ( c-p_1(t)-p_2(t)+\rho_I(t)x_I \right ) \right ] \\
&~~~~~~~~~~~~~~+  \gamma_I g_I^*(t,x_I) \int_{\R^+}g_I^*[t,x_I-(y-l_1-l_2)]-g_I^*(t,x_I)v(\sd y)\bigg\}\\
=&\sup_{ \{l_1,l_2\}\in \mathscr{A}_I}\left \{\mathscr{L}^{l_1,l_2,\xi_1,\xi_2}_I[V_I^*(t,x_I)]-\frac{\gamma_I}{2}\int_{\R^+}\big[g_I^*(t,x_I-(y-l_1-l_2))-g_I^*(t,x_I)\big]^2v(\sd v)\right \}\\
=&\sup_{\{l_1,l_2\}\in \mathscr{A}_I}\bigg\{\frac{\partial V_I^*(t,x_I)}{\partial t}+\frac{\partial V_I^*(t,x_I)}{\partial x_I}\left ( \int_{\R^+}[y -l_1-l_2+\theta y-\xi_1(t)l_1^2-\xi_2(t)l_2]v(\sd y)+\rho_I(t)x_I\right ) \\
&~~~~~~~~~~~~~~ +\int_{\R^+}V_I^*[t,x_I-(y-l_1-l_2)]-V_I^*(t,x_I)v(\sd y)-\frac{\gamma_I}{2}\int_{\R^+}\big[g_I^*(t,x_I-(y-l_1-l_2))-g_I^*(t,x_I)\big]^2v(\sd v)\bigg\}.\\
\end{align*}
Since we can write $V_I^*(t,x_I)=e^{\int_t^T \rho_I(s) \d s} x_I + B_I(t)$ and $g_I^*(t,x_I)=e^{\int_t^T \rho_I(s) \d s} x_I + b_I(t)$, by substituting them into the equation above, we obtain
\begin{align}\label{proof3.2:eq1}
 &\mathrm{HJB}_I (t, x_I ;l_1, l_2 , \xi_1, \xi_2  )\notag \\
  =&\sup_{\{l_1,l_2\}\in \mathscr{A}_I}\bigg\{-\rho(t)e^{\int_t^T\rho_I(s)\d s}x_I+B_I'(t)
+e^{\int_t^T\rho_I(s) \d s}\Big[\int_{\R^+}[y -l_1-l_2+\theta y-\xi_1(t)l_1^2-\xi_2(t)l_2]v(\sd y) +\rho_I(t)x_I\Big] \notag \\
&-\int_{\R^+}[ e^{\int_t^T\rho_I(s) \d s}(y-l_1-l_2)]v( \sd y)-\frac{\gamma_I}{2}\int_{\R^+}e^{2\int_t^T\rho_I(s) \d s}(y-l_1-l_2)^2v(\sd y)\bigg\} \notag \\
=&B_I'(t)
+e^{\int_t^T\rho_I(s)\d s}\sup_{\{l_1,l_2\}\in \mathscr{A}_I}\bigg\{\int_{\R^+}\theta y-\xi_1(t)l_1^2-\xi_2(t)l_2-\frac{\gamma_I}{2}e^{\int_t^T\rho_I(s) \d s}(y-l_1-l_2)^2v( \sd y)\bigg\}.
\end{align}
By the first-order condition, we have
\begin{equation} \label{ineqset}
\begin{cases}
-2\xi_1(t)l_1+\gamma_I e^{\int_t^T \rho_	I(s) \d s}(y-l_1-l_2)=0,\\
-\xi_2(t)+\gamma_Ie^{\int_t^T \rho_	I(s) \d s}(y-l_1-l_2)=0,
\end{cases}
\end{equation}
which further implies
\begin{align*}
 l_1 = \frac{\gamma_I e^{\int_t^T \rho_I (s) \d s}}{2 \xi_1(t) + \gamma_I e^{\int_t^T \rho_I(s) \d s}} (y-l_2) =y-l_2-\frac{\xi_2(t)}{\gamma_I e^{\int_t^T \rho_I \d s}}
\end{align*}
Recall that $l_1 \in [0, y]$, $l_2 \in [0,y]$ and $l_1 + l_2 \in [0, y]$. If $y > \frac{\xi_2(t)}{\gamma_I e^{\int_t^T \rho(s)\sd s}}+\frac{\xi_2(t)}{2\xi_1(t)} $, the two first-order conditions can be satisfied by
\begin{align*}
\begin{cases}
l_1 = \frac{\xi_2(t)}{2\xi_1(t)},\\
l_2 = y - \frac{\xi_2(t)}{\xi_1(t)} - \frac{\xi_2(t)}{\gamma_I e^{\int_t^T} \rho_I(s)\d s}.
\end{cases}
\end{align*}
If $y \le\frac{\xi_2(t)}{\gamma_I e^{\int_t^T \rho(s)\d s}}+\frac{\xi_2(t)}{2\xi_1(t)} $, the
\begin{align*}
\begin{cases}
l_1 = \frac{\gamma_Ie^{\int_t^T \rho_I (s) \d s}}{2 \xi_1(t)+ \gamma_I e^{\int_t^T} \rho_I(s)\d s} y,\\
l_2 = 0,
\end{cases}
\end{align*}
leads to the supreme in \eqref{proof3.2:eq1}.

Combining the two solutions above, we have the equilibrium reinsurance strategies associated with $\xi_1$ and $\xi_2$ as
\begin{equation*}
l_1^*(t, y;\xi,\xi_2)=\begin{cases}
q(t)y,& y\leq d(t)\\
\frac{\xi_2(t)}{2\xi_1(t)},& y>d(t)
\end{cases},
~~~\text{ and }~~
l_2^*(t, y;\xi_1,\xi_2)=\begin{cases}
0,& y\leq d(t)\\
y-d(t),& y>d(t)
\end{cases}
\end{equation*}
%
where $q(t)=\frac{\gamma_I e^{\int_t^T \rho_I(s) \d s}}{2\xi_1(t)+\gamma_I e^{\int_t^T \rho_I(s) \d s }}$ and $d(t)=\frac{\xi_2(t)}{\gamma_I e^{\int_t^T \rho(s) \d s}}+\frac{\xi_2(t)}{2\xi_1(t)}$.

\end{proof}
\subsection{Proof of Lemma \ref{ghG}}
\begin{proof}
First note that
\begin{align*}
e(x)=&\frac{2}{x^2}\sum_{n=1}^{\infty}\frac{1}{n!}x^n- \left (1+\frac{2}{x}+\frac{2}{x^2}\right )
=\frac{2}{x^2}+\frac{2}{x}+1+\frac{2}{x^2}\sum_{n=3}^{\infty}\frac{1}{n!}x^{n}- \left (1+\frac{2}{x}+\frac{2}{x^2} \right)
=\sum_{n=1}^{\infty}\frac{2}{(n+2)!}x^{n}.
\end{align*}
Obviously, $e(x)$ is strictly increasing in $x$ for positive $x$. Thus, we have $\lim_{x \to 0} e(x)=0$, $\lim_{x \to +\infty} e(x)=+\infty$. In short, $e:(0,+\infty) \to (0,+\infty)$ is a continuous and strictly increasing function.

The monotonicity of $g(x)$ can be obtained by determining its derivative as follows:
{\small
\begin{align*}
2g'(x)=&-(2C_{R1}+1)+\frac{2[2C_{R1}-1+x(2C_{R1}+1)](2C_{R1}+1)+8C_{R1}}{2}\left[ \left (2C_{R1}-1+x(2C_{R1}+1)\right) ^2+8C_{R1}(x+1)\right]^{-\frac{1}{2}}\\
=&(2C_{R1}+1) \left ( -1+\frac{ 2C_{R1}-1+x(2C_{R1}+1)+\frac{4C_{R1}}{(2C_{R1}+1)}}{\left ( [2C_{R1}-1+x(2C_{R1}+1)]^2+8C_{R1}(x+1)\right )^{\frac{1}{2}}} \right ) \\
=&(2C_{R1}+1) \left ( -1+\left ( \frac{[2C_{R1}-1+x(2C_{R1}+1)]^2+[\frac{4C_{R1}}{(2C_{R1}+1)}]^2+2[2C_{R1}-1+x(2C_{R1}+1)]\frac{4C_{R1}}{(2C_{R1}+1)}}{[2C_{R1}-1+x(2C_{R1}+1)]^2+8C_{R1}(x+1)}\right )^{\frac{1}{2}} \right ) \\
=&(2C_{R1}+1) \left ( -1+\left ( \frac{[2C_{R1}-1+x(2C_{R1}+1)]^2+8C_{R1}x+[\frac{4C_{R1}}{(2C_{R1}+1)}]^2+\frac{8C_{R1}(2C_{R1}-1)}{2C_{R1}+1}}{[2C_{R1}-1+x(2C_{R1}+1)]^2+8C_{R1}x+8C_{R1}}\right )^{\frac{1}{2}}\right )\\
=&(2C_{R1}+1)\left ( -1+\left ( \frac{[2C_{R1}-1+x(2C_{R1}+1)]^2+8C_{R1}x+8C_{R1}\bigg[[\frac{2C_{R1}}{(2C_{R1}+1)^2}]+\frac{(2C_{R1}-1)}{2C_{R1}+1}\bigg]}{[2C_{R1}-1+x(2C_{R1}+1)]^2+8C_{R1}x+8C_{R1}}\right )^{\frac{1}{2}} \right ) \\
=&(2C_{R1}+1) \left (-1+\left ( \frac{[2C_{R1}-1+x(2C_{R1}+1)]^2+8C_{R1}x+8C_{R1}\bigg[\frac{4C_{R1}^2+2C_{R1}-1}{(2C_{R1}+1)^2}\bigg]}{[2C_{R1}-1+x(2C_{R1}+1)]^2+8C_{R1}x+8C_{R1}}\right )^{\frac{1}{2}}\right )\\
=&(2C_{R1}+1) \left ( -1+\left ( \frac{[2C_{R1}-1+x(2C_{R1}+1)]^2+8C_{R1}x+8C_{R1}\bigg[1-\frac{2C_{R1}+2}{(2C_{R1}+1)^2}\bigg]}{[2C_{R1}-1+x(2C_{R1}+1)]^2+8C_{R1}x+8C_{R1}}\right )^{\frac{1}{2}} \right )
\end{align*}
}
Because $\frac{2C_{R1}+2}{(2C_{R1}+1)^2}>0$, we have
\[\frac{[2C_{R1}-1+x(2C_{R1}+1)]^2+8C_{R1}x+8C_{R1}\bigg[1-\frac{2C_{R1}+2}{(2C_{R1}+1)^2}\bigg]}{[2C_{R1}-1+x(2C_{R1}+1)]^2+8C_{R1}x+8C_{R1}\big]}<1, ~~~\text{ for } x > 0 .\]
This implies that $g'(x)<0$. Furthermore, we have
$$\lim_{x \to 0} g(x)=\frac{-(2C_{R1}-1)+\sqrt{(2C_{R1}-1)^2+8C_{R1}}}{2}~~~\text{ and }~~ \lim_{x \to +\infty} g(x)=\frac{2C_{R1}}{2C_{R1}+1} . $$ Thus, $g:(0,+\infty) \to \left(\frac{2C_{R1}}{2C_{R1}+1},\left[-(2C_{R1}-1)+\sqrt{(2C_{R1}-1)^2+8C_{R1}}\right]/2\right)$ is continuous and strictly decreasing.

Combining the monotonicity and continuity of both $ e(\cdot ) $ and $ g(\cdot) $, we obtain that $G(x):=g(e(x))$ is a continuous and strictly decreasing function.
\end{proof}

\subsection{Proof of Proposition \ref{exist}}
\begin{proof}
From the definition
$$
h_1(\xi_1)=\frac{\gamma_Ie^{\int_t^T\rho_I(s)\d s} G^{-1}(\frac{\gamma_{R1}e^{\int_t^T\rho_{R1}(s) \d s}}{\xi_1})}{ \beta \left ( 1+\frac{\gamma_Ie^{\int_t^T\rho_I(s)\d s} }{2\xi_1} \right ) } ~~\text{ and }~~
h_2(\xi_1)= \frac{\gamma_Ie^{\int_t^T\rho_I(s)\d s}}{\beta}\left ( 1+C_{R2}-\frac{\gamma_Ie^{\int_t^T\rho_I(s)\d s}}{\gamma_Ie^{\int_t^T\rho_I(s)\d s }+2\xi_1}\right ),
$$
we have
\[
\begin{aligned}
h_1(\xi_1)-h_2(\xi_1)
=&\frac{\gamma_Ie^{\int_t^T\rho_I(s) \d s}}{\beta}\left ( \frac{ G^{-1}(\frac{\gamma_{R1}e^{\int_t^T\rho_{R1}(s)\d s}}{\xi_1})}{ 1+\frac{\gamma_Ie^{\int_t^T\rho_I(s) \d s} }{2\xi_1} }-  ( 1+C_{R2} ) +\frac{\frac{\gamma_Ie^{\int_t^T\rho_I(s) \d s}}{2\xi_1}}{\frac{\gamma_Ie^{\int_t^T\rho_I(s) \d s}}{2\xi_1}+1} \right ) \\
=&\frac{\gamma_Ie^{\int_t^T\rho_I(s) \d s}}{ \beta \left ( 1+\frac{\gamma_Ie^{\int_t^T\rho_I(s) \d s} }{2\xi_1} \right ) }\left ( G^{-1}(\frac{\gamma_{R1}e^{\int_t^T\rho_{R1}(s) \d s}}{\xi_1})- (1+C_{R2}) \left ( 1+\frac{\gamma_Ie^{\int_t^T\rho_I(s) \d s} }{2\xi_1}\right )+\frac{\gamma_Ie^{\int_t^T\rho_I(s) \d s}}{2\xi_1}\right )\\
=&\frac{\gamma_Ie^{\int_t^T\rho_I(s) \d s}}{ \beta \left ( 1+\frac{\gamma_Ie^{\int_t^T\rho_I(s) \d s} }{2\xi_1} \right ) }\left ( G^{-1}(\frac{\gamma_{R1}e^{\int_t^T\rho_{R1}(s) \d s}}{\xi_1})-C_{R2}\frac{\gamma_Ie^{\int_t^T\rho_I(s) \d s}}{2\xi_1}-(1+C_{R2})\right ) .
\end{aligned}
\]
It is known from Lemma \ref{ghG} that $G(\cdot)$ is a continuous and strictly decreasing function, and
\begin{align*}
     \lim_{x \to 0} G(x)& =\lim_{x \to 0} g(e(x))=\lim_{x \to 0} g(x)=\frac{-(2C_{R1}-1)+\sqrt{(2C_{R1}-1)^2+8C_{R1}}}{2} , \\
     \lim_{x \to +\infty} G(x) &=\lim_{x \to +\infty} g(e(x)) =\lim_{x \to +\infty}  g(x) =\frac{2C_{R1}}{2C_{R1}+1}.
\end{align*}
Thus, $G^{-1}:(\frac{2C_{R1}}{2C_{R1}+1},\frac{-(2C_{R1}-1)+\sqrt{(2C_{R1}-1)^2+8C_{R1}}}{2}) \to (0,+\infty)$ is also a continuous and strictly decreasing function.


As $\xi_1 \to \frac{\gamma_{R1}e^{\int_t^T\rho_{R1}(s) \d s}(2C_{R1}+1)}{2C_{R1}}$, we have $\frac{\gamma_{R1}e^{\int_t^T\rho_{R1}(s) \d s}}{\xi_1} \to \frac{2C_{R1}}{2C_{R1}+1}$ and $\frac{\gamma_Ie^{\int_t^T\rho_I(s)\sd s}}{2\xi_1} \to \frac{1}{2C_{R1}+1}$. Thus,
$$
G^{-1}(\frac{\gamma_{R1}e^{\int_t^T\rho_{R1}(s)\d s}}{\xi_1})-C_{R2}\frac{\gamma_Ie^{\int_t^T\rho_I(s) \d s}}{2\xi_1}\rightarrow \infty.
$$

As $\xi_1 \to \frac{2\gamma_{R1}e^{\int_t^T\rho_{R1}(s) \d s}}{\sqrt{(2C_{R1}-1)^2+8C_{R1}}-(2C_{R1}-1)}$, we have $\frac{\gamma_{R1}e^{\int_t^T\rho_{R1}(s)\d s}}{\xi_1} \to \frac{\sqrt{(2C_{R1}-1)^2+8C_{R1}}-(2C_{R1}-1)}{2}$ and $\frac{\gamma_Ie^{\int_t^T\rho_I(s) \d s}}{2\xi_1} \to \frac{\sqrt{(2C_{R1}-1)^2+8C_{R1}}-(2C_{R1}-1)}{4C_{R1}}$. Thus,
$$G^{-1}(\frac{\gamma_{R1}e^{\int_t^T\rho_{R1}(s) \d s}}{\xi_1})-C_{R2}\frac{\gamma_Ie^{\int_t^T\rho_I(s) \d s}}{2\xi_1}
\rightarrow -\frac{\sqrt{(2C_{R1}-1)^2+8C_{R1}}-(2C_{R1}-1)}{4C_{R1}}C_{R2}<0.
$$

%

Together with the facts that $ 1 +C_{R2} > 0 $ and $ h_1(\xi_1) - h_2 ( \xi_2 ) $ is a continuously decreasing function on its domain, we conclude that $h_1(\xi_1) - h_2 ( \xi_2 )  $ crosses the $x$-axis from above on the interval $ (\frac{2\gamma_{R1}e^{\int_t^T\rho_{R1}(s) \d s}}{\sqrt{(2C_{R1}-1)^2+8C_{R1}}-(2C_{R1}-1)}, \, \frac{\gamma_{R1}e^{\int_t^T\rho_{R1}(s) \d s}(2C_{R1}+1)}{2C_{R1}}) $.
Therefore, there exists $ \bar{\xi}_1$ such that $h_1(\bar{\xi}_1)-h_2(\bar{\xi}_2)$, i.e., $h_1(\bar{\xi}_1)=h_2(\bar{\xi}_2)=\bar{\xi}_2$.
\end{proof}

\end{appendices}

\bibliography{refer}
\bibliographystyle{apacite}

\end{document}